# Scientific Prizes and the Extraordinary Growth of Scientific Topics

**Authors:** Ching Jin[1,2], Yifang Ma[1,3], Brian Uzzi*[1,2]

**Affiliations:**

[1]Northwestern Institute on Complex Systems (NICO), Northwestern University, Evanston IL, 60208, USA.
[2]Kellogg School of Management, Northwestern University, Evanston IL, 60208, USA.
[3]Department of Statistics and Data Science, Southern University of Science and Technology, Shenzhen, Guangdong 518055, China
*Corresponding Author: uzzi@northwestern.edu

**Abstract**

Fast growing scientific topics have famously been key harbingers of the new frontiers of science, yet, large-scale analyses of their genesis and impact are rare. We investigate one possible factor connected with a topic's extraordinary growth: scientific prizes. Our longitudinal analysis of nearly all recognized prizes worldwide and over 11,000 scientific topics from 19 disciplines indicates that topics associated with a scientific prize experience extraordinary growth in productivity, impact, and new entrants. Relative to matched non-prizewinning topics, prizewinning topics produce 40% more papers and 33% more citations, retain 55% more scientists, and gain 37% and 47% more new entrants and star scientists, respectively, in the first five-to-ten years after the prize. Funding do not account for a prizewinning topic's growth. Rather, growth is positively related to the degree to which the prize is discipline-specific, conferred for recent research, or has prize money. These findings reveal new dynamics behind scientific innovation and investment.

Introduction

The extraordinary growth of a scientific topic occurs when there is a period of unexpected and abnormally large growth in a topic's impact or size[1-3]. Classical studies of extraordinary growth focus on the specifics of illustrative cases, such as the Copernican revolution and its associated changes in astronomical topics' productivity and impact on scientific thinking[1,4]. Newly-available, large-scale data on nearly all scientific topics enables the study of periods of extraordinary growth from a statistical perspective. Here we investigate generalizable factors associated with the onset and magnitude of the extraordinary growth of scientific topics using a large and diverse sample of scientific topics[2,5-7].

We investigated the statistical dynamics around a possible correlate of the onset of a scientific topic's extraordinary growth: a topic's association with a scientific prize[8,9]. Scientific prizes were originally restricted to certain scientific disciplines and topics, but they have proliferated and are now awarded in nearly all disciplines[3,9]. Prize research has primarily studied how awards change prizewinners' careers[10-16] by recognizing[17,18], celebrating[9,18,19], and changing perceptions of a scholar's work[2,20,21]. For example, case studies of Howard Hughes Medical Investigators and John Bates Clark and Fields medalists have shown that prizewinners' papers published before their prize is conferred gain citations significantly faster than expected[10-12,14,15,22] after the prize and that winning one prize increases the probability of the same scholar winning future prizes[9].

It is uncertain whether the link between prizes and unexpected growth for a single prizewinner's work extends to changes in the growth of an entire topic, and current theoretical arguments and empirical work are nascent. On the one hand, scholars have argued that the increased interest in a prizewinning scientist's work may likewise increase interest in the topic associated with a prize[11,23,24]. On the other hand, scholars have claimed that prizes signal that the best work associated with a topic has been done, which would lessen rather than an expand interest in a topic[11,25]. An impression reinforced by the fact that prizewinners tend to move onto new topics after winning a prize[10,12] and the papers by matched contenders for the prize are concommitently cited less than expected[11,12].

As a first step toward understanding the possible links between prizes and a topic's extraordinary growth, we investigated whether prizes are statistically associated with the onset of a period of abnormal growth in a topic's productivity, impact, and the migration of scientists into and out of the topic. Our analysis uses new data on hundreds of recognized scientific prizes worldwide[9] and longitudinal data on over 10 thousand scientific topics.

From various sources, we collected data on 405 scientific prizes conferred 2,900 times between 1970 and 2007 with respect to 11,539 scientific topics in 19 disciplines. Scientific prize data was collected from Wikipedia pages on prizes and scientists. The prizes in our sample include awards like the Wolf Prize and Turing Prize, as well as hundreds of other prizes recognized on Wikipedia's "scientific prizes" page[9,16]. To validate the Wikipedia data, we manually cross-checked it with prize-related data on dedicated webpages and in print media. Fig. 1a illustrates how prizes and topics are linked. Prizes were linked to topics by associating the prizewinning scientist with the topics to which they are considered to have made meaningful contributions, also known as their "known-for" topics. We defined known-for topics empirically and via crowdsourcing. Empirically, known-for topics are the topics on which a scientist has published 10 or more total papers. We cross-validated this operationalization using Wikipedia's "Known-For" dataset, which lists a scientist's known-for topics based on the crowdsourced opinions of scientists and Wikipedia users (see Methods section for details). Consistent with our 10-paper threshold, we found that a scholar's Wikipedia "Known-For" topics were those topics scholars had 10 papers on average (See Supplementary Fig. S1 for alternative thresholds and robustness checks).

Scientific topic data comes from Microsoft Academic Graph (MAG). MAG covers over 172 million publications by 209 million authors in 48,000 journals from 1800 to 2017. It uses crowdsourcing, AI, and NLP to generate meaningful and plausible classifications of scientific topics in ways that integrate human expert observations and empirical replicability. MAG defines the universe of scientific topics using Wikipedia article pages that are classified as being on "scientific topics"; these pages are created and updated by scientists and users through crowdsourcing. MAG then uses NLP and AI to assign research publications to topics by associating a paper's text (not just keywords) with the linguistic content of Wikipedia's scientific topics pages[26]. As a validity check, MAG uses statistical sampling to verify the face validity of the classification of papers to topics[27,28] (see SI. Sec.1 for details).

To test whether a prizewinning topic's growth after the prize is awarded is significantly greater than expected, i.e., "extraordinary," we used a difference-in-differences (DID) regression design. DID regression tests for the effects of randomly assigned treatments as well as non-causal tests of statistical significance, in which one group experiences an event that the other group does not experience[11-13,29]. Following prior prize research[11-13], we used DID to test for non-causal statistical relationships between prizewinning and a topic's extraordinary growth. Also, following the literature, we combined DID with Dynamic Optical Matching (DOM)[30-33]. DOM is a longitudinal case control matching procedure that identifies a matched group of non-prizewinning topics. (The Method section reports measurement details and the formal regression equation.)

Using DOM[30-33], we identified five non-prizewinning topics that had growth patterns statistically equivalent to prizewinning topics before the prize year on six growth criteria. Three growth criteria broadly represented a topic's impact: (a) productivity, (b) number of citations, and (c) number citations of the topic's leading scientists. Three criteria represent the movement of scientists into and out of a topic: (a) number of incumbent scientists, (b) number of new entrants, and (c) number of star scientists from outside the topic[26,34-38]. To ensure that prizewinning topics and their matched topics have yearly parallel growth before the prize year, prizewinning and matched topics had to be from the same discipline and had to have statistically indistinguishable growth patterns on all six measures, each year, for 10 years before the prize[39,40].

"Productivity" measures a topic's yearly number of publications and relates to a topic's output, resources, and publication norms[41]. "Citations" measures a topic's yearly number of citations and captures a topic's impact[42]. Citation growth was measured on an absolute and per capita basis. "Impact of Topic's Leading Scientists" measures the mean total citations of the upper 5% of scientists working on a topic and captures the impact of a topic's intellectual leaders. To measure a topic's size[36], we created the variable "#Incumbents," which measures a topic's yearly number of continuing scientists, and the variable "#Entrants," which measures the yearly number of scientists who publish for the first time on a topic. "#Disciplinary Stars" measures the number of scientists working on a topic that are among the 5% most cited scientists in the topic's discipline[43]. The last three variables ensure that matched topics and prizewinning topics have equivalently eminent scientists and likely prizewinners[12,44]. Supplementary Fig. S4 demonstrates that prizewinning and matched topics have no statistical differences on all matching variables prior to the prize year (11 years* 6 measures = 66 tests, all p-values > 0.2).

To test for a statistical relationship between prizes and extraordinary growth, we compared the post-prize growth in prizewinning topics at time *t* against the average growth rate of the matched topics at time *t*. We refer to this difference in growth as $\Delta_t$, formally expressed as:

$$\Delta_t = \log(Y_t) - \log(\widetilde{Y}_t), \qquad (1)$$

$Y_t$ is the prizewinning topic's growth at time $t$ and $\widetilde{Y}_t$ is the same quantity based on the average growth of the matched topic group (see Fig. 1b for an illustration). Time $t$'s range is -10 to +10. A $\Delta_t = 0.0$ indicates no difference in the growth of the prizewinning and matched groups. Time *t* represents actual time in years (i.e., no rescaling). Extraordinary growth occurs when $\Delta_t$ is statistically and significantly different from zero. We analyze the growth dynamics associated with a topic's first prize and consider a prize to be correlated with the onset of extraordinary growth if growth starts shortly after the prize year. The SI reports confirmatory robustness checks for alternative measures of $\Delta_t$ (Supplementary Fig. S2).

Results

Prizewinning is strongly and positively related to and extraordinary growth. Relative to matched topics, prizewinning topics have unexpected and significant increases in growth on all six measures of growth. Table 1 shows the DID regression results for all six growth measures net of controls. Prizewinning ($\beta_1$) is a binary variable equal to one or zero for prizewinning and matched topics, respectively. The estimated $\beta_1$s all have insignificant p-values greater than 0.05, indicating that prizewinning and matched topics have no statistically significant growth differences before the prizewinning event on all six measures of growth. Post ($\beta_2$) is a binary variable equal to one for the 10-year period after the prize year and equal to zero before the prize year. Prizewinning*post ($\beta_3$) is the interaction term. $\beta_2$ and $\beta_3$ are significant for all six growth measures (all p-values <0.001), demonstrating that prizewinning topics grow unexpectedly larger after the prize is awarded relative to matched topics.

Figures 2a-f plot the average magnitudes of extraordinary growth $\Delta_t$ ($e^{\Delta t} - 1 = (Y_t - \tilde{Y}_t)/\tilde{Y}_t$) for all six growth measures. Notably, extraordinary growth begins the year following the prize and continues for at least the next 10 years. At five years after the prize, prizewinning topics have grown an average 17% to 30% larger than matched topics depending on the growth measure ($\Delta_5$, all p-values <0.0001). At 10 years after the prize, the growth gap increases to 25% to 55% depending on the growth measure ($\Delta_{10}$, all p-values<0.0001).

In comparing the overall impact of prizewinning topics to matched topics, we observe large and consistent growth differences. Prizewinning topics display strong post-prize extraordinary growth in productivity and citation impact. At 10 years after the prize, prizewinning topics are 39.8% more productive in terms of the number of publications ($\Delta_{10} = 0.3351$, $e^{\Delta_{10}} - 1 = 0.3981$) than matched topics (Fig. 2a). Figure 2b and 2c show changes in citation impact. Prizewinning topics have 32.6% more yearly citations ($\Delta_{10}=0.2825$, $e^{\Delta_{10}} - 1 = 0.3264$) and experience a 7.75% average increase in per capita citations per paper at year 10 over matched topics (Supplementary Fig. S3), indicating that the increase in citation impact holds on a per scientist basis as well. Lastly, the citation impact of leading scientists within the topic is 25% greater than the impact of leading scientists working on non-prizewinning matched topics ($\Delta_{10} = 0.2232$, $e^{\Delta_{10}} - 1 = 0.2500$, Fig 2c).

The migration of scientists into and out of a topic is an indicator of a topic's research attractiveness. Prizewinning topics retain significantly more incumbents (Fig. 2d). After the prize year, incumbent scientists continue to routinely publish on that topic at a rate that is 54.8% ($\Delta_{10} = 0.4366$, $e^{\Delta_{10}} - 1 = 0.5475$) higher than incumbent scientists publishing in matched topics.

New entrants enter prizewinning topics in abnormally high numbers. Prizewinning topics gain over 36.7% more new entrants on average than do matched topics ($\Delta_{10} = 0.3129$, $e^{\Delta_{10}} - 1 = 0.3673$, Fig. 2e). About half of new entrants (46.3%) are rookie scientists who make their first publication on the prizewinning topic. This indicates that prizewinning topics attract young scholars who often make longer-term research commitments[45].

Star scientists are at the other end of their career stage relative to rookie scientists. Star scientists are the 5% most highly-cited scholars in their discipline (physics, chemistry, sociology, etc.)[46]. Star scientists move into prizewinning topics in larger numbers than into matched topics.

Counting star scientists working on prizewinning and matched topics before and after the prize, we found that prize prizewinning topics gain over 47% more star scientists from across the whole discipline than do matched topics ($\Delta_{10} = 0.3878, e^{\Delta_{10}} - 1 = 0.4737$, Fig. 2f).

Finally, we tested whether extraordinary growth is associated with a topic's paradigmatic diversification, a relationship hypothesized in the literature[1]. We created a master list of the topics that new entrants into prizewinning and matched topic groups had published before the prize year. We used Shannon Entropy to quantify the topic diversity of the lists (see SI). The findings suggest that extraordinary growth positively correlates with paradigmatic diversification. As the $\Delta_{10}$ of topics increase, the paradigmatic diversification of the prizewinning topic grows systematically greater than the paradigmatic diversification of the matched topic group. For example, for topics with a $\Delta_{10}$ value of 1.5, the prizewinning topic group's paradigmatic diversity is 11.6% greater than the matched topic group (see Supplementary Fig. S7 for all change comparisons).

Funding and Extraordinary Growth

Does funding explain the results? Funding can provide resources that may affect a topic's growth and impact[47]. Also, funding agencies explicitly ration their funds to the most promising research topics[48,49], which implies that topics that receive grants are seen by funders as having special growth potential. To examine whether funding differences between prizewinning and matched topics explain the extraordinary growth of prizewinning topics, we collected funding data on the subset of our data for which it is available. The NIH publishes funding data in the form of a public list of all papers funded by a specific grant. Using this list, we found that a subsample of 2,853 out of the 11,539 prizewinning topics in our full sample received NIH funding from 1985 to 2005. For the sample of 2,853 topics, we used DOM to create a new matched group of five non-prizewinning topics for each prizewinning topic, just as we did in the main analysis to ensure that the prizewinning and matched topic groups were balanced and met the parallel trends criteria. In this subsample, we used the same six matching criteria as in the main sample. Seventy-six percent of the matched topics turned out to also be NIH grant recipients, which makes sense, since funding should correlate with our six matching criteria.

To further validate the preceding analysis, we conducted a second analysis with a different subsample. In the second subsample, we matched prizewinning NIH funded topics based on the six criteria used in the main analysis plus a seventh criteria: the matched topics had to have received NIH funding. The results replicate the funding analysis shown above and the main findings, further demonstrating that funding differences do not explain the extraordinary growth of prizewinning topics (please see Supplementary Information, Supplementary Fig. S8, Tab. S13).

Based on the above findings, we concluded that research funding appears to be uncorrelated with a topic's extraordinary growth, at least within this subsample of NIH projects. First, prizewinning topics have equivalent or slightly less NIH funding before the prize than matched topics (Fig. 3a-c). Second, the level of NIH funding of prizewinning topics is the same before and after prize year (Fig. 3d). Third, consistent with the main results, after accounting for funding levels of prizewinning and matched topics, we find that the onset of the extraordinary growth of prizewinning topics begins the first year after the prize year (Fig. 3e-j).

Topic-by-Topic Generalizability and Placebo Tests

Our main findings compare differences in post-prize growth for prizewinning and matched topic groups. Here, we examined the growth of each prizewinning topics ($N = 11,539$) and their five matched topics separately. After the prize, 60% of the prizewinning topics show growth larger than that of their specific five matched topics (binomial test, all p-values<0.001). This finding indicates that the results are not driven by outlier topics and are generalizable (See Supplementary Tab. S1). The binomial tests were reinforced by placebo tests[40,50,51]. The placebo test examines whether the matched topics showed abnormal growth following the prize year of the prizewinning topic (even if they grew slower than prizewinning topics). Supplementary Fig. S5 shows that matched topics have no coincidental extraordinary growth (all p-values>0.05), reinforcing our main finding—prizewinning is associated with the onset of a topic's sustained period of unexpected and extraordinary growth.

Prize Characteristics and the Post-Prize Magnitude of Extraordinary Growth

A prize's characteristics and a topic's growth dynamics may be interrelated.[9,24] For example, prizes have been explicitly established for specific fields or with money to send "a symbolic message to the general public that perhaps science and scientists really mattered [24]". We researched three features of prizes that appear broadly in prize data: money, discipline-specificity, and research recency [18,19].

Moneyed prizes provide symbolic cultural value and dedicate valuable resources to a topic[19,25]. For example, the 2012 Fundamental Physics Prize, 2013 Tang Prize in Chinese Studies, and the Breakthrough Prize in the Life Science are monied prizes created to communicate to scientists and the public that "the best minds should make at least as much as any trader on Wall Street" and that science aims to "contribute to world development[52]." Nevertheless, the impact of monetary prizes is unknown. On the one hand, prize money might raise perceptions of a topic's importance, and on the other hand, extrinsic rewards can reduce intrinsic motivation or make little topic-wide difference if money goes to a single individual, not to science[52]. We operationalized prize money as a binary — money vs. no money variable (45% of prizes were coded as monied). We also created a three-category money variable defined as (a) no money, (b) money below the median, and (c) money above the median (see Supplementary Tab. S11), which confirmed the simpler binary variable reported below.

Discipline-specific prizes often have greater perceived within-discipline status than do general prizes. For instance, the Fields Medal in math is generally viewed as more prestigious than the general-science National Medal of Science Prize[18,19]. The discipline-specific variable takes on the value of one if at least 85% of all the winners of the prize come from the same discipline; zero otherwise (78% of prizes were coded as field-specific).

Though prizes can be given for recent or past research, most are generally awarded for contemporary work, not longstanding research; otherwise, many prizes would exclude younger scholars[53]. For example, the MacArthur Prize website acknowledges that the fellowship "…is not a reward for past accomplishment, but rather an investment in a person's originality, insight, and potential." To code a prize for its association with recent or past research, we counted the interevent time in years between a prize year and the first year a prizewinner worked on the prizewinning topic. We then created a distribution of inter-event times across all prizes.

To examine the link between prize characteristics and magnitude of extraordinary growth, we regressed the $\Delta_{10}$ of our six growth variables on money, discipline-specificity, and recency, along with control variables. Control variables include lagged values of each growth trend at times *t*-1, *t*-2, and *t*-3 years to account for autoregressive effects of $\Delta_{10}$. To account for differences in a prize's visibility, we added control variables for the prize's average Wikipedia page views during 2017, the number of past conferrals of a prize up to the prize year, and the prize's age at the prize year. To control for multiple prize conferrals in the same year, we added a binary variable for whether a topic had multiple prize recipients. To control for the prizewinner's scientific status, we added a binary variable for whether the prizewinner is a star scientist, a designation measured as being among a topic's top 5% of cited authors. Fixed effects for discipline and year control differences that vary with the discipline and year of prizewinning. Supplementary Information presents details on variable measurements and descriptive statistics.

Figure 4 plots the raw data relationship between a prize's features and $\Delta_{10}$ for our six growth variables. Prize characteristics significantly predict the magnitude of $\Delta_{10}$ in 17 out of 18 cases. The 17 positive tests are statistically significant and sometimes substantively large. The only null relationship out of the 18 tests indicates that prize money does not predict changes in the citation impact of a topic's leading scientists.

The significant relationship between prize characteristics and $\Delta_{10}$ generalizes when controlling for confounds. Fig. 4d-f report the standardized coefficients for all the predictor variables net of control variables. Of the three prize features, recency and specialization have the largest substantive associations with $\Delta_{10}$. For example, conditional on the topic being a prizewinner, a single standard deviation increase in recency is associated with a 13.8% increase in $\Delta_{10}$ of new scientists and a 14.6% increase in $\Delta_{10}$ of citations. Moneyed or field-specific prizes predict an increase in $\Delta_{10}$ of 1.9% or 5.3% in citation growth, respectively. These findings indicate that prizes are associated with extraordinary growth in a topic and that the magnitude of extraordinary growth is strongly predicted by a prize's features. Supplementary Tables S2-S10 report all regression-detailed estimates, robustness checks, and other fit statistics[54].

Discussion

The rapid, unexpected growth of scientific topics have long held a reputation in science because of their connection to unexpected shifts in research efforts, technology developments, and individual scientific careers. Yet, most work has focused on intricate analyses of specific cases rather than on the statistical dynamics of extraordinarly growth. We conducted a large-scale, science-wide analysis of one factor that may be associated with the onset and magnitude of growth: scientific prizes. We found that scientific prizes predict unexpectedly large changes in the growth of a research topic.

Relative to matched, non-prizewinning topics with equivalent historical growth, disciplinary status, and demography, prizewinning topics are significantly more productive, higher impact, and attractive to incumbent, rookie, and star scientists. At five-to-10 years after the prize year, prizewinning topics are over 30% more productive and are associated with significantly higher citation impact on average on a per capita basis. Also, we found that prize characteristics predict the magnitude of a topic's unexpected and extraordinary growth. The magnitude of growth is greatest when the topic of research associated with the prize is recent, the prize is a discipline-specific rather than general science award, and the award is monied. Surprisingly, we found no evidence that funding is related to the extraordinary growth of prizewinning topics. These findings have implications for the study of the mechanisms behind the extraordinary growth of prizewinning scientific topics.

Possible Mechanisms Behind the Extraordinary Growth of Scientific Topics

While our work is among the first to look at how prizewinning relates to the growth of ideas rather than the careers of individual scientists, the mechanisms by which prizewinning plays a role in the abnormal growth of topics remains an important avenue of future research. We focused on the statistical links between prizes and the onset of extraordinary growth, which can provide a basis for examining mechanisms. Our findings relate to what Thomas Kuhn called the "essential tension between tradition and innovation" [1,3,55]. Kuhn argued that conservative investment and risk-taking are interconnected in scientific advances. Consistent with Kuhn, we found that prizewinning topics retain a base of incumbents who perhaps reinvest in the topic with a conservative orientation that takes current ideas further, while at the same time the new entrants attracted to prizewinning topics bring with them new viewpoints that promote a risk-taking orientation for experimentation.

The new points of view that entrants bring to the prizewinning topic may also be connected to Kuhn's notion of paradigm shifts. Paradigm shifts have been identified in a few detailed case studies and focus on a topic's revolutionary growth in size and impact. For example, Kuhn references work on the Copernican revolution that indicated that the earth-centric view of the solar system was replaced with sun-centric view of the solar system over a more than 100-year period. Our work suggests that a possible early source of a paradigm shift maybe the onset of a topic's unexpected growth, which involves an influx of new scientists whose research expertise and diverse demographics provide the potential building blocks of new paradigms, a notion compatible with the teams, social network, and organizational literatures that have shown that new entrants with diverse backgrounds can transform thinking and stimulate innovation [56-61].

Our findings also relate to Zuckerman's prizes-as-signals argument[62]. In this theoretical argument, Zuckerman argued that prizes may act as signals to scientists that a prizewinning topic offers comparatively strong prospects for professional growth, or conversely, prizes signal the end of a period of growth by suggesting that a topic's potential opportunities for professional recognition have run their course, leading scientists to move away from the prizewinning topic[25].

Examining our findings in light of these crisscrossing arguments suggests conditional support for the claim that a prizewinning topic's intellectual and professional attractiveness become more rather than less positive after the prize's conferment. First, we found that the prize characteristics of money, disciplinary-specialization, and recency are positively related to the onset and magnitude of extraordinary growth. Second, we found that rookie and star scientists begin working on prizewinning topics in unexpectedly high numbers after the prize's award. Nevertheless, a firm conclusion may have to wait for new data or emerging natural experiments that could shed more definitive light on these causal arguments and their contingencies.

Funding stands out as a possible driver of the growth of topics because funding decisions intentionally involve peer and expert reviews designed to steer researchers and resources to topics predicted to be promising research areas[47,48,63]. For example, special funding was dedicated boost rates of interdisciplinary research[57,64,65]. Regardless of the stated objectives of funding, Myers (2020) found that funding weakly induces scientists to change their research direction[47].

Consistent with Myers' work, we found that available data on prizewinning and non-prizewinning topics do not differ in NIH grants received before the prize's conferral, and funding remains flat for prizewinning topics after the prize, indicating that funding levels do not explain the association between prizes and a topic's extraordinary growth. While this result is consistent with other research showing that funding does not materially change research agendas already in process[47], the finding does rely only on NIH funding, which is a large and leading source of funding, but does not necessarily generalize to all types of funding or funding practices. Nevertheless, with the number of topics growing each year[59] while funding levels have remain relatively flat suggests that the link between funding and extraordinary growth may be an especially important target for future research pursuits.

Future Prize Research

A central question in the scientific prize literature since its inception has been whether prizes "select" the highest quality research[9,62]. To date, the answer to this question has remained beyond the reach of empirical work, including in this work, as the answer depends on many factors such as the multiple ways in which an idea's quality can be defined and measured. What has been determined is that prizes are awarded on average to worthy work. For example, the official statement by the International Congress of Mathematicians, which include the Field's Medal, noted that on average prizes are not intended to select the best work, just work that is worthy: "We must bear in mind how clearly hindsight shows that past recipients of the Field's medal were only a selection from a much larger group of mathematicians whose impact on mathematics was at least as great as that of the chosen" (ICM 1994).[12]

If one accepts the ICM's approach to prizes, newly identified linkages among prizewinners and prizes raise related questions about prizes and meritocracy. Ma et al. (2018) found that the number of prizes worldwide have proliferated, which presumably offers more opportunity to celebrate scientists and ideas[9]. However, as prizes have proliferated, prizes have become increasingly concentrated within a relatively small network of elite scientists and their mentees. Further, the most successful mentees study topics that differ from their mentor's topics of study, suggesting that skills associated with conducting prizewinning work can be mentored and applied to new research areas[16]. Future research may therefore examine ways in which prizewinning tacit knowledge and research skills can be diffused more broadly across scientists and topics, and how vacancy chains within the prize network may open and close.

Another area of prize meritocracy focuses on representation. New research indicates that women are winning prizes more inline with their participation rates in science but continue to be but statistically less likely to win research prizes than their male peers[18]. Focusing on research areas where women's participation is large and there are many prizes, such as biomedicine, women are winning prizes a rates closer to their representation in the field but the prizes they win have lower prestige and less money than the prizes won by their male counterparts[18]. Those trends coupled with the finding that a prize's prestige and money predict the magnitude of extraordinary growth a prize can have on a topic suggests that future research should investigate how prize committee nomation and decision procedures are linked to diversity, a process already taking place among some prize committees[66].

Relatedly, the question of prize meritocracy in science depends on the availability of data that can connect scientists, topics of study, and impact. Here we used the relatively new Microsoft Academic Graph (MAG) data, which uses expert human opinion and algorthims to identify differences in ideas and topics. The relatively new benefits of the MAG data is that it has explicit reproducbile procedures that can be evaluated for bias. Nevertheless, science is an ever changing entity. Year to year, new fields emerge and the semantic meaning of terms change over time and at different rates across different communities of practice, which means MAG provides a defensible snapshot of scientific topics, but not necessarily an immutable one.

In this way, our work contributes to knowledge about changes in scientific thinking, the mobility of scientists between topics, and investment funding that together broaden understanding of how the frontiers of science develop.

Methods

Difference-in-Difference Regression

To study whether a growth pattern before and after a topic is associated with a topic's prizewinning event, we combine difference-in-difference regression with matching. This approach has been used frequently in studies of prizes and in other areas where observation data is not randomly assigned and where experimental random assignment may be impossible due to factors such as harm to subjects[12,22,29]. Borjas and Doran[12] studied how Field Medal prizewinners' career change relative to non-prizewinners. In their case, like ours, prizes were not assumed to be randomly assigned to topic. Hence, they used matching, where matching is done between the prizewinner and a set of "contenders" non-prizewinners whose historical scholarly performance would have made them contenders for the prize. To justify matches between prizewinners and non-prizewinners, Borjas and Doran stated, "In short, while it is tempting to claim that the 52 Fields Medal recipients are in a class by themselves and that there are no losing contenders with equivalent or better early achievements, this view does not correspond with what mathematicians themselves have written.[12]" For example, The International Congress of Mathematicians[67], which awards the Fields Medal, the Nevanlinna Prize, the Gauss Prize, and the Chern Medal during the congress's opening ceremony has officially stated: "We must bear in mind how clearly hindsight shows that past recipients of the Fields Medal were only a selection from a much larger group of mathematicians whose impact on mathematics was at least as great as that of the chosen" (ICM 1994)[12]. The arbitrariness in the number, timing, and field distribution of Fields Medal recipients means that a similarly great group of "contenders" should exist that can be contrasted with the winners in a difference-in-differences strategy to determine how winning the medal influences productivity and research choices." Similarly, the DID regression used in our analysis is specified as follows:

$$Z_{i,t} = \beta_0 + \beta_1 prizewinning_i + \beta_2 post_t + \beta_3 prizewinning_i * post_t + fixed\ effect + \epsilon_{i,t}, \quad (2)$$

where $Z_{i,t}$ is one of our six outcome variables for topic $i$ at time $t$. $Prizewinning_i$ is a dummy variable quantifying whether the topic $i$ is a prizewinning topic or a non-prizewinning topic from a matched group. $Post_t$ is a dummy variable measuring whether time $t$ is before or after the prizewinning event[68]. If the topic belongs to the matched group, the prizewinning year of the related prizewinning topic will be used as the reference point. $\epsilon_{i,t}$ is an error term. Regression is specified with fixed effects for discipline and prize year and robust standard errors. The parallel trend assumption (PTA) is essential for DID analysis. We met the PTA criteria by matching prizewinning and non-prizewinning topics on six separate growth indices that uses a procedure that ensures that the pre-prize growth indices of the prizewinning and matched group parallel each other (Fig. 2). Moreover, our analysis meets the PTA criteria of "indistinguishability." "Indistinguishability" is a stronger version of the PTA criteria. PTA requires growth is constant. Indistinguishability requires the growth trends are constant and have zero difference. In our analysis, both groups have constant growth and have zero difference from each other.

Dynamic Optimal Matching Procedure

To select the matched non-prizewinning topics, we used a Dynamic Optimal Matching method, which applies the Optimal Matching Method[30,32,33] to a time-series data to simultaneously maximize the closeness and balance characteristics of accurate matching[30,31,33,69]. We focused on prizewinning between 1970 and 2007, resulting in a sample of 12,041 prizewinning and peer non-prizewinning topics. 11,539 topics (> 95%) could be statistically matched to non-prizewinning topics using DOM. If a topic has multiple prizes over its lifetime, we match for the first prizewinning. We also ensure that the selected known-for topic has been studied by the prizewinner before the associated prizewinning event. Since we want to study the relationship between prizewinning and topic's growth pattern, we need 10 years of data prior to the prizewinning event to perform the matching method, and 10 years after prizewinning to allow different statistical analyses of the growth pattern. Hence, we measure the growth pattern of topics from 1960 to 2017.

First, we selected a matched topic candidate pool[32]. To achieve this, for each prizewinning topic $i$, we selected up to 40 close-distance topics in terms of a distance measure ($\theta_{i,j}$) from the same discipline, generating a peer non-prizewinning candidate pool. For 95% of the prizewinning topics, a proper non-prizewinning candidate pool was identified (11,539/12,041= 95.8%). To achieve matching, we defined a distance measure $\theta_{i,j}$ to quantify the closeness between the prizewinning topic $i$ and a non-prizewinning topic $j$[29,70]:

$$\theta_{i,j} = \frac{\sum_{n=1}^{N} \sum_{t=t^*-t_0}^{t^*} (\log Y_{i,n}(t) - \log Y_{j,n}(t))^2}{N * (t_0 + 1)} \quad (3),$$

where $Y_{i,n}$ indicates the quantity for the topic $i$ in terms of one of the $N = 6$ matched categories (i.e. Productivity, Citations, Lead Scientist impact, #incumbents, #Entrants and #Disciplinary Stars). $t$ measures number of years prior to the prizewinning year for topic $i$, where $t^*$ represents the first prizewinning year for topic $i$, and $t_0 = 10$, indicates we traced the growth pattern for topics in an 11-year duration, including 10 years prior to the prizewinning year. To account for possible correlations among different variables, we repeated the analyses using an alternative distance measure -- Mahalanobis distance -- to quantify the closeness of the topics, finding consistent results (Supplementary Fig. S6).

Second, to ensure the balance between the matched and prizewinning topics for the entire system, we select 5 matching topics from the candidate pool to be the topic's peer group. In this process, we (1) minimized the distances between the matched and prizewinning topics in terms of $\theta_{i,j}$, and (2) made sure the distribution of the matched and prizewinning topics are acceptably and simultaneously close enough for all 66 covariates.

Specifically, we make sure that the differences between the prizewinning and matched topic groups are small enough for each matching category $n$ and for any time $t$ before the prizewinning event, where the differences between the prizewinning topic $i$ and its expected growth at time $t$ and category $n$ are quantified by $\Delta_{i,n}(t) = (\log Y_{i,n}(t) - \log \tilde{Y}_{i,n}(t))$. The expected growth is obtained by averaging the trajectory of the matched topics. This problem is a classical optimization problem, which could be solved with typical Mixed Integer Programming (MIP) methods[31,32]. We found the best-optimized matching possible where (1) the distance between the prizewinning topics and the matched topics is minimized; at the same time, (2) the difference between the matched and prizewinning groups is not statistically significant for any $t \in [-10,0]$ and $n \in [1,6]$. Mathematically, we have:

$$\left| \frac{\sum_{i=1}^{M} \Delta_{i,n}(t)}{M} - 0 \right| < 1.96 * SE\left(\Delta_{i,n}(t)\right). \quad for \ \forall \ 1 \leq n \leq 6, -10 \leq t \leq 0. \quad (4)$$

Here $SE(\Delta_{i,n}(t))$ measures the standard error of the $\Delta_{i,n}(t)$ for the prizewinning topics at time $t$ and in category $n$, and $M$ captures number of prizewinning topics. To prevent outlier bias from topics with a large $\Delta_{i,n}(t)$ in the MIP process, we also measured the topic-by-topic growth of each individual topic. Specifically, for any $t$ ($-10 \leq t \leq 0$), we compare the growth pattern of each prizewinning topic and all of its five matched topics. For each of the 66 covariances, we ensure prizewinning topics had equal probability to show larger or smaller growth than their matched topics. This method not only guarantees closeness between the matched and prizewinning topics but also unsure good balancing between and within topic groups. The method has also been validated by an additional placebo test. Please see Supplementary Information and Supplementary Fig. S4-S6 for detailed information.

Acknowledgments

We thank Jonas Lindholm-Uzzi, Meghan Stagl, and Yiting Wen for their help in data collection. We thank Dashun Wang, Benjamin F. Jones, Nima Dehmami, Yang Yang, Yuan Tian, Lu Liu, Junming Huang, Yi Bu, Sourav Medya, Yian Yin, Liqiang Huang, Zhixiang Ma, Mohammad Rassolinejad for helpful discussions. Funding: This material is based upon work supported wholly or in part by the Northwestern Institution on Complex Systems (NICO) and the Air Force Office of Scientific Research under award number FA9550-19-1-0354.Author Contributions: CJ, YM, and BU collected data, designed the study, and wrote the paper. CJ and BU conducted analyses. Competing Interests: The authors declare no competing interests in this work. Author Contributions: CJ, YM, and BU collected data, designed the study, and wrote the paper. CJ and BU conducted analyses. Competing Interests: The authors declare no competing interests in this work. Data Availability: Data used in this work are publicly available from Wikipedia and Microsoft Academic Graph (MAG). The MAG dataset could be accessed through API: https://www.microsoft.com/en-us/research/project/microsoft-academic-graph/. The Wikipedia information could be obtained: https://en.wikipedia.org/wiki/Main_Page. Code Availability: The code used in this study is available from the corresponding author upon reasonable request.

**DID Regression of a Topic's Comparative Post-Prizewinning Growth on Six Measures**

| Growth Measures | (1) | (2) | (3) | (4) | (5) | (6) |
|---|---|---|---|---|---|---|
| | **Productivity** | **Citations** | **Impact of Topic's Lead Scientists** | **#Incumbents** | **#Entrants** | **#Disciplinay Stars working on the topic** |
| Prizewinning ($\beta_1$) | 0.005 | -0.002 | 0.002 | 0.003 | -0.005 | 0.004 |
| S.E. | (0.0387) | (0.0442) | (0.0259) | (0.0408) | (0.0413) | (0.0362) |
| *p*-value | (0.8979) | (0.965) | (0.9295) | (0.9444) | (0.9042) | (0.9176) |
| Post ($\beta_2$) | 0.609*** | 1.240*** | 1.047*** | 0.973*** | 0.750*** | 0.667*** |
| S.E. | (0.0115) | (0.0151) | (0.00970) | (0.0135) | (0.0117) | (0.0119) |
| *p*-value | (0.0000) | (0.0000) | (0.0000) | (0.0000) | (0.0000) | (0.0000) |
| Prizewinning * Post | 0.213*** | 0.169*** | 0.188*** | 0.271*** | 0.206*** | 0.254*** |
| S.E. | (0.0134) | (0.0172) | (0.0128) | (0.0157) | (0.0138) | (0.0137) |
| *p*-value | (8.207e-57) | (8.735e-23) | (3.447e-48) | (2.928e-66) | (2.126e-50) | (9.208e-76) |
| Fixed Effect Controls: | | | | | | |
| Discipline | Yes | Yes | Yes | Yes | Yes | Yes |
| Year | Yes | Yes | Yes | Yes | Yes | Yes |
| const | 2.926*** | 4.502*** | 4.237*** | 1.136*** | 3.217*** | 1.358*** |
| S.E. | (0.123) | (0.136) | (0.0988) | (0.135) | (0.131) | (0.117) |
| *p*-value | (2.860e-124) | (3.571e-236) | (0) | (3.475e-17) | (7.454e-131) | (7.269e-31) |
| *N* | 1,453,914 | 1,453,914 | 1,453,914 | 1,453,914 | 1,453,914 | 1,453,914 |
| *R*-sq | 0.154 | 0.301 | 0.392 | 0.290 | 0.209 | 0.274 |

Standard errors in parentheses. * $p<0.05$, ** $p<0.01$, *** $p<0.001$

**Table 1: Prizewinning topics are associated with extraordinary growth.** $Prizewinning_i$ is a dummy variable quantifying whether the topic $i$ is a prizewinning topic or a non-prizewinning topic from the matched groups. $Post_t$ is a dummy variable measuring whether time $t$ is before or after the prizewinning event. If the topic belongs to the matched group, the prizewinning year of the related prizewinning topic will be used as the reference point. Regression is specified with fixed effects for discipline and prize year and robust standard errors. All estimated $\beta_1$s have p-values > 0.05, indicating that prizewinning and matched topics have no differences before the prizewinning event. $\beta_3$s are significant for all six categories (all p-values <0.001 even after adjustments were made for multiple comparisons), demonstrating that prizewinning topics grow unexpectedly larger than matched topics after the year the prize is conferred. OLS models are used to perform the DID analysis.

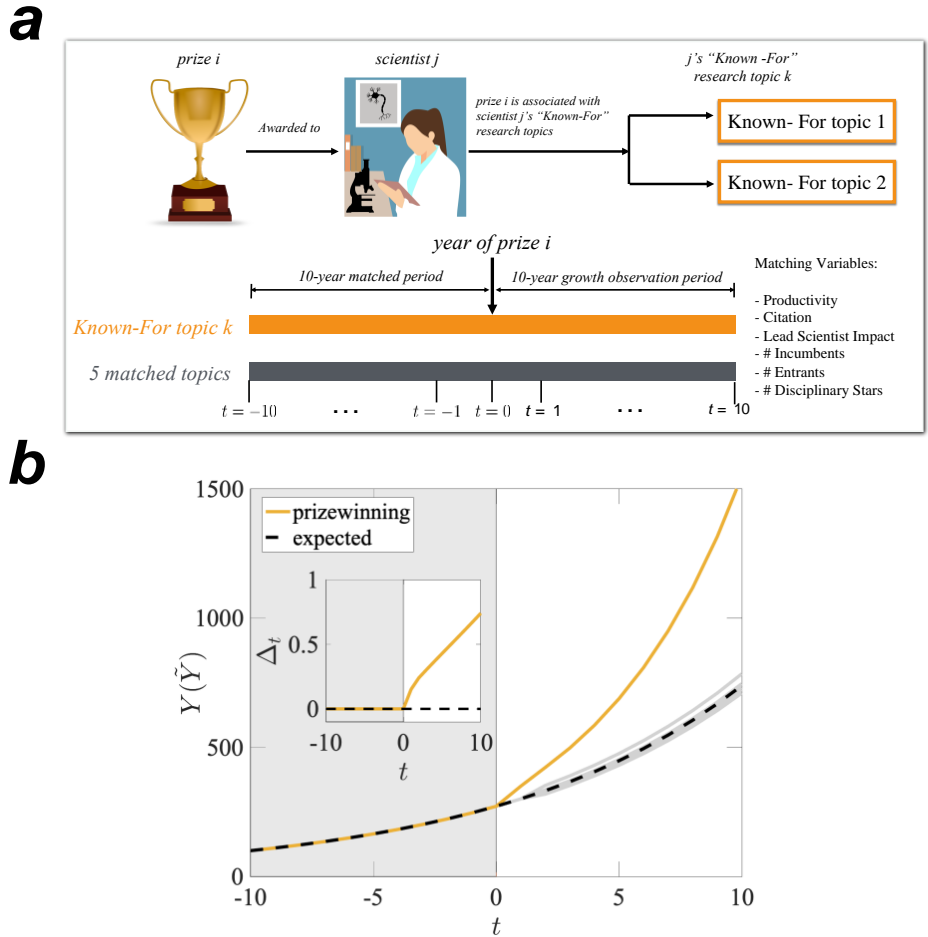

**Figure 1: Procedure for defining prizewinning topics.** **a)** Our procedure has three steps. First, we collected a large sample of over 400 recognized prizes recorded in Wikipedia. The Wikipedia sample includes universally known prizes (e.g., Wolf prize, Kyoto, Lasker etc.) and prizes accredited by associations some of which are less well-known to the public or scientific community but significant for association members. Second, we linked prizes to prizewinners using the prize's homepage and the prizewinner's Wikipedia page. Third, topics are linked to prizes via the prizewinner's Known For (KF) topics. KF topics are the small subset of topics that a researcher is known for out of all the topics a researcher has worked on. We defined a researcher's KF topics as those topics on which the researcher had published 10 papers or more, which was cross-validated with Wikipedia's KF topics pages, which uses crowdsourcing from scientists and users to identify a researcher's KF topics (Robustness Check, Supplementary Fig. S1). Each prizewinning topic was then matched with five non-prizewinning topics that had statistically indistinguishable year-to-year growth from the prizewinning topic during the 10-year period prior to the prizewinning year. Matched topics were from the same discipline as the prizewinning topic. Six measures of growth were used in the matching: productivity, citations, impact of a topic's lead scientists, number of incumbent scientists, number of new scientists, and number of disciplinary star scientists. Illustration photos are obtained from www.pxfuel.com and www.pikpng. com. **b)** $\Delta_t$ measures the difference between the log of the average growth of prizewinning topics minus the log of the average growth of matched topics after the prize year. The inset shows $\Delta_t$ as the relative percentage growth of the prizewinning topic compared to the matched topic group, hence the time prior to prize year, $t=0$ is flat. $\Delta_t$ is measured and reported separately for 6 growth measures of a topic as described in the text. Gray panel represents the time before the prize year.

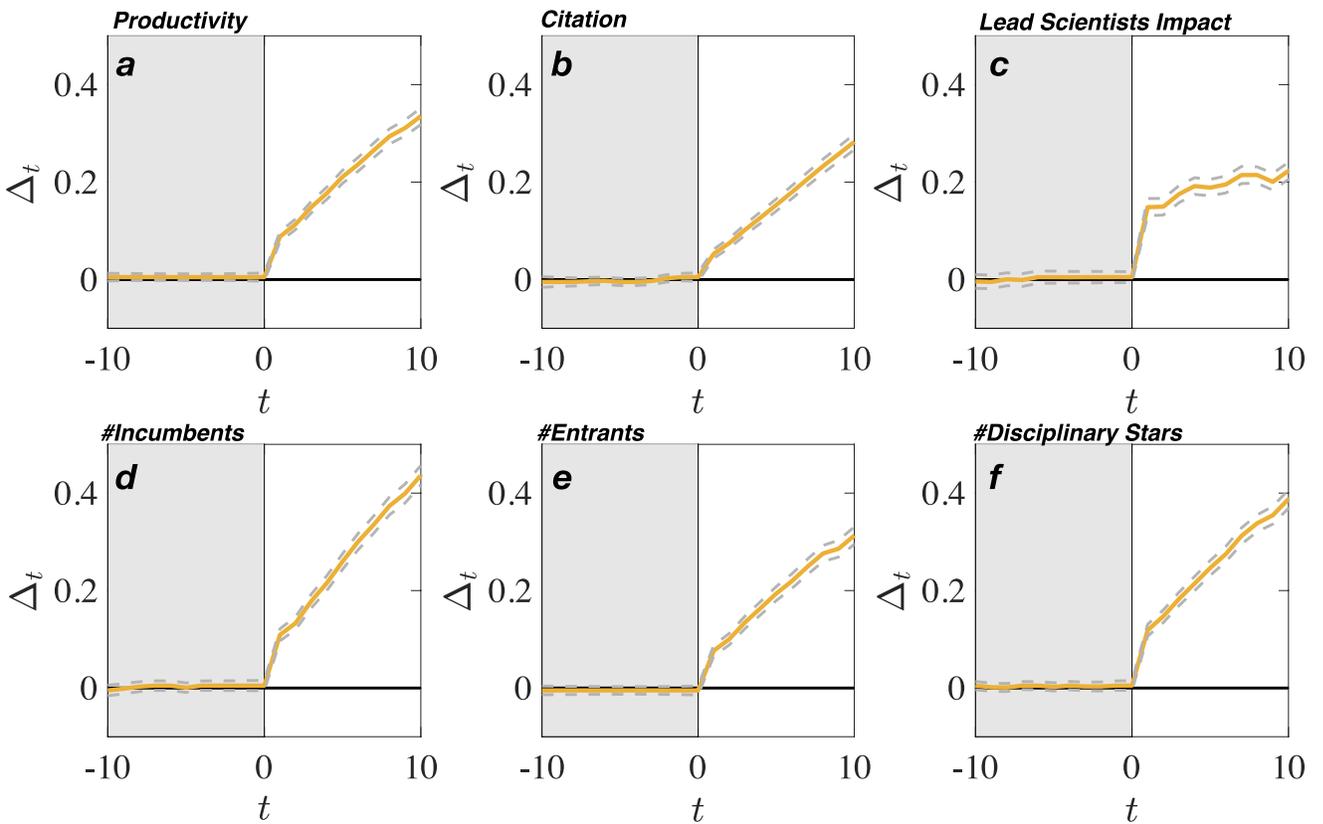

**Figure 2. Scientific Prizes and Extraordinary Growth.** Panels (a) through (f) show the differences in growth rates of prizewinning and matched topics for 10 years prior to and 10 years after the prize year in relation to (a) productivity, (b) citation, (c) topic's lead scientist impact, (d) #incumbent scientists, (e) #entrants, and (f) #disciplinary stars. Statistically significant growth differentials between prizewinning (gold line) and matched topics (flat black line) begin shortly after the prize (vertical line) and continue yearly following the prize (95% CIs shown as dashed lines). At 10 years, the growth rates of prizewinning topics exceed matched topics by 25% to 55% depending on the growth variable (two tailed t-test, p<0.001. *p*-value ranges from 0 to $1.56 * 10^{-139}$). Prizewinning and non-prizewinning matched topic groups had not growth differences for 10 years prior to the prize (two-tailed t-test, *p*>0.05 for all 6*11=66 tests, *p*-values range from 0.1438 to 0.9310).

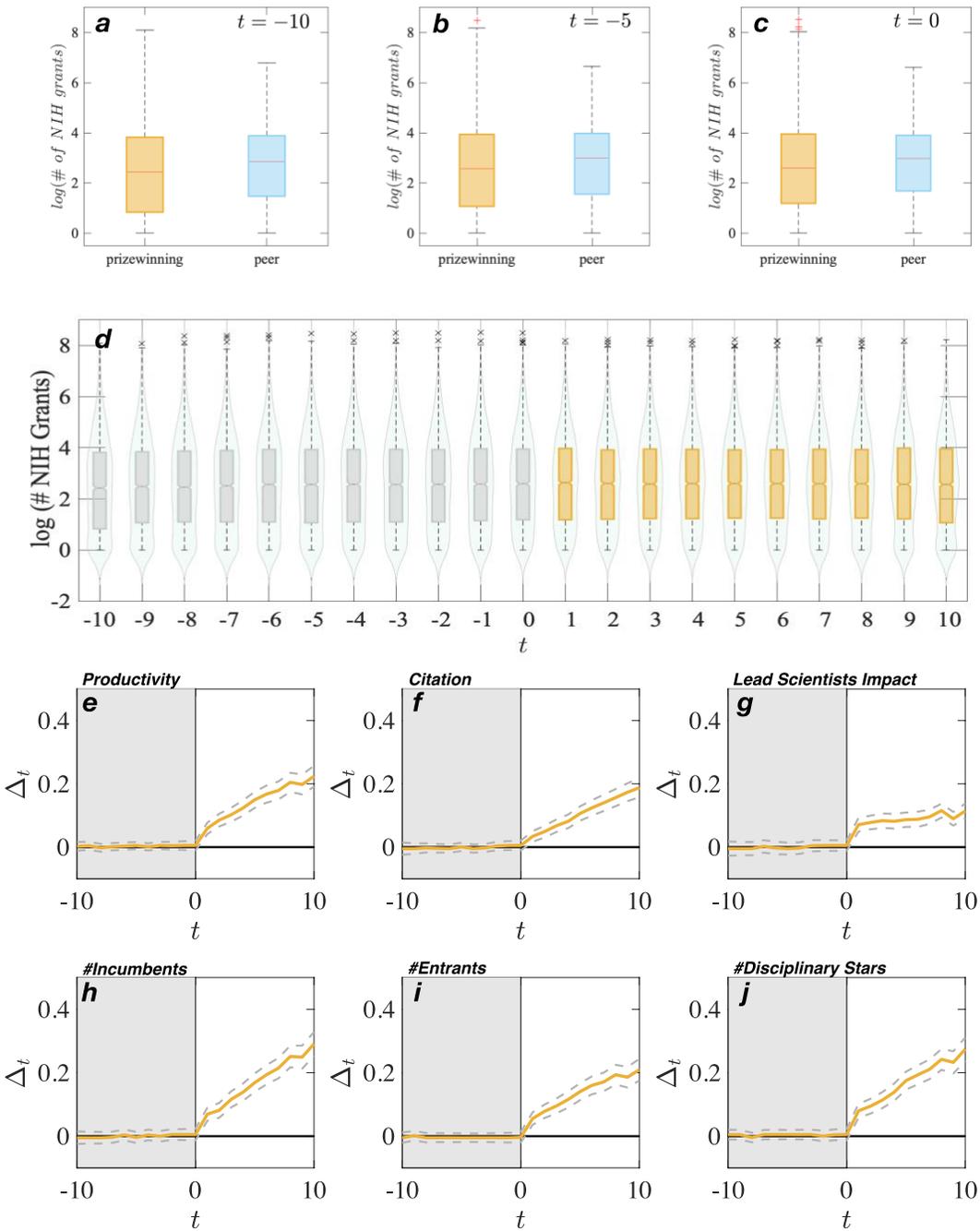

**Figure 3. Funding and Extraordinary Growth are Statistically Unrelated in NIH Subsample.** The analysis examines whether funding plays a role in the extraordinary growth of prizewinning topics. In the analysis, we find a subset of 2,853 prizewinning topics that were matched on six measures of growth and the receipt of NIH funding, which is publicly available. Repeating the main analysis that was done on the full sample of over 11,000 topics, we found that funding does not explain the extraordinary growth of prizewinning topics. First, (a-c) shows that prizewinning have statistically equivalent or less funding before the prize than matched topics. Second, (d) shows that NIH grant funding is largely flat before (grey box) and after (gold box) the prizewinning event for the prizewinning topics. The center line of the box plot is the median of the normalized grants, box limits correspond to the data's first and third quartiles, notches represent 95% CI, and violin plots represent the data's distribution. Third, plots e-j show that prizewinning topics grow relatively larger than expected after the prize year consistent with the main analysis with the full sample.

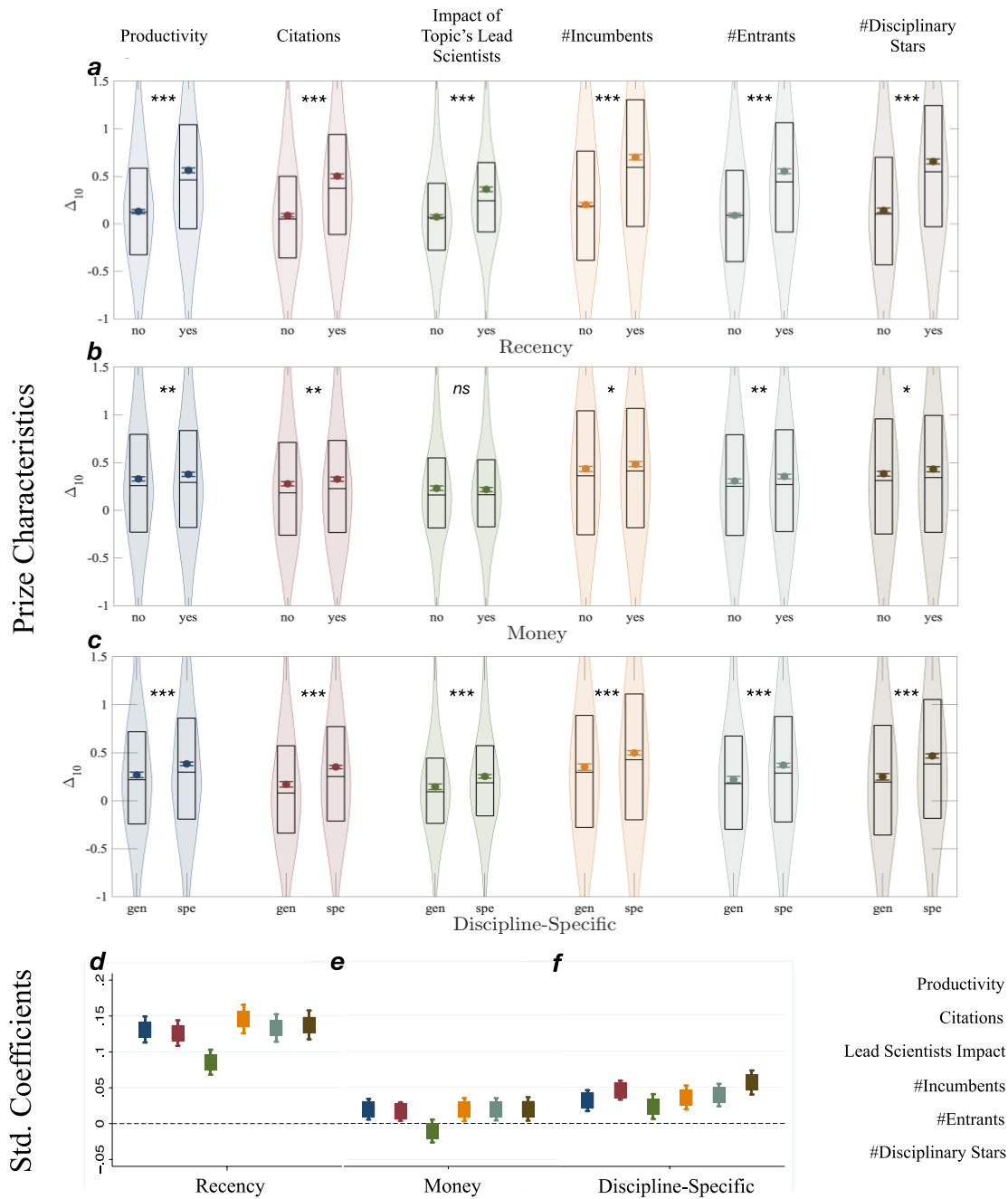

**Figure 4. Prize Characteristics Predict the Magnitude of Extraordinary Growth.** (a) through (c) show the raw data relationship between a prize's characteristics and the estimated magnitude of $\Delta_{10}$ on 6 growth measures (*** $p<0.001$; ** $p<0.01$; and * $p<0.05$). The color dots with error bars represent the mean value and 95% CI, the center line of the black box plot is the median, box limits correspond to the data's first and third quartiles, and violin shades represent the data's distribution. Two-sided t-test are used in these analyses. The *p*-value for (a) ranges from $2.70*10^{-180}$ to $5.30*10^{-67}$, for (c) ranges from $1.60*10^{-25}$ to $9.06*10^{-9}$, and for (b) *p*-value ranges from 0.0017 to 0.013 except for lead scientist impact (*p*-

value =0.374). Sample sizes for topics associated with prizes with recency, money and discipline specific are n=5,672 (46.73%), n=6,102 (50.27%) and n=8,812 (72.59%) respectively. (d)-(f) show the standardized coefficient from a regression of $\Delta_{10}$ on all three prize characteristics, indicating the significant and substantive association between prize characteristics and the magnitude of the expected level of extraordinary growth in a topic after it is associated with a prize. Control variables are lagged $\Delta_t$ at $t$ = -1, -2, -3, discipline, year, prize visibility and reputation, and prizewinner status. Error bars demonstrate 95% CI. Please see Supplementary Tab. S2-S10 for details of regression analyses and further robustness checks.

## References

1  Kuhn, T. S. *The Structure of Scientific Revolutions*.  (University of Chicago, 1970).
2  Mazloumian, A., Eom, Y.-H., Helbing, D., Lozano, S. & Fortunato, S. How citation boosts promote scientific paradigm shifts and nobel prizes. *PloS one* **6**, e18975 (2011).
3  Foster, J. G., Rzhetsky, A. & Evans, J. A. Tradition and innovation in scientists' research strategies. *American Sociological Review* **80**, 875-908 (2015).
4  Nickles, T. in *The Stanford Encyclopedia of Philosophy*  (ed Edward N. Zalta) (Metaphysics Research Lab, Stanford University, https://plato.stanford.edu/archives/win2017/entries/scientific-revolutions/, 2017).
5  Shapin, S. *The scientific revolution*.  (University of Chicago Press, 2018).
6  Kornmesser, S. Scientific revolutions without paradigm-replacement and the coexistence of competing paradigms: The case of generative grammar and construction grammar. *Journal for General Philosophy of Science* **45**, 91-118 (2014).
7  Gillies, D. in *Heuristic reasoning*    89-112 (Springer, 2015).
8  Zuckerman, H. *Scientific Elite: Nobel Laureates in the United States*.  (Free Press, 1977).
9  Ma, Y. & Uzzi, B. Scientific prize network predicts who pushes the boundaries of science. *Proceedings of the National Academy of Sciences* **115**, 12608-12615 (2018).
10 Azoulay, P., Graff Zivin, J. S. & Manso, G. Incentives and creativity: evidence from the academic life sciences. *The RAND Journal of Economics* **42**, 527-554 (2011).
11 Reschke, B. P., Azoulay, P. & Stuart, T. E. Status spillovers: The effect of status-conferring prizes on the allocation of attention. *Administrative Science Quarterly*, 0001839217731997 (2017).
12 Borjas, G. J. & Doran, K. B. Prizes and productivity how winning the fields medal affects scientific output. *Journal of human resources* **50**, 728-758 (2015).
13 Bricongne, J.-C. Do prizes in economics affect productivity? *Science Po Publications* **24** (2014).
14 Chan, H. F., Gleeson, L. & Torgler, B. Awards before and after the Nobel Prize: A Matthew effect and/or a ticket to one's own funeral? *Research Evaluation* **23**, 210-220 (2014).
15 Chan, H. F., Frey, B. S., Gallus, J. & Torgler, B. Does the John Bates Clark Medal boost subsequent productivity and citation success?  (2013).
16 Ma, Y., Mukherjee, S. & Uzzi, B. Mentorship and protégé success in STEM fields. *Proceedings of the National Academy of Sciences* (2020).
17 Lincoln, A. E., Pincus, S., Koster, J. B. & Leboy, P. S. The Matilda Effect in science: Awards and prizes in the US, 1990s and 2000s. *Social studies of science* **42**, 307-320 (2012).
18 Ma, Y., Oliveira, D. F. M., Woodruff, T. K. & Uzzi, B.   (Nature Publishing Group, 2019).

19  English, J. F. *The economy of prestige: Prizes, awards, and the circulation of cultural value*. (Harvard University Press, 2008).
20  Stiglitz, J. Give prizes not patents. *New Scientist* **16** (2006).
21  Moser, P. & Nicholas, T. Prizes, Publicity and Patents: Non-Monetary Awards as a Mechanism to Encourage Innovation. *The Journal of Industrial Economics* **61**, 763-788 (2013).
22  Chan, H. F., Frey, B. S., Gallus, J. & Torgler, B. Academic honors and performance. *Labour Economics* **31**, 188-204 (2014).
23  Azoulay, P., Liu, C. C. & Stuart, T. E. Social influence given (partially) deliberate matching: Career imprints in the creation of academic entrepreneurs. *American Journal of Sociology* **122**, 1223-1271 (2017).
24  Zuckerman, H. Views: The Sociology of the Nobel Prize: Further Notes and Queries: How successful are the Prizes in recognizing scientific excellence? *American scientist* **66**, 420-425 (1978).
25  Zuckerman, H. The proliferation of prizes: Nobel complements and Nobel surrogates in the reward system of science. *Theoretical Medicine* **13**, 217-231 (1992).
26  Griffiths, T. L. & Steyvers, M. Finding scientific topics. *Proceedings of the National Academy of Sciences* **101**, 5228-5235, doi:10.1073/pnas.0307752101 (2004).
27  Sinha, A. *et al.* in *Proceedings of the 24th international conference on world wide web.* 243-246 (ACM).
28  Wang, K. *et al.* A Review of Microsoft Academic Services for Science of Science Studies. *Frontiers in Big Data* **2**, 45 (2019).
29  Jin, G. Z., Jones, B., Lu, S. F. & Uzzi, B. The reverse Matthew effect: Consequences of retraction in scientific teams. *Review of Economics and Statistics* **101**, 492-506 (2019).
30  Rosenbaum, P. R. Optimal matching for observational studies. *Journal of the American Statistical Association* **84**, 1024-1032 (1989).
31  Zubizarreta, J. R. Using mixed integer programming for matching in an observational study of kidney failure after surgery. *Journal of the American Statistical Association* **107**, 1360-1371 (2012).
32  Pimentel, S. D., Kelz, R. R., Silber, J. H. & Rosenbaum, P. R. Large, sparse optimal matching with refined covariate balance in an observational study of the health outcomes produced by new surgeons. *Journal of the American Statistical Association* **110**, 515-527 (2015).
33  Rosenbaum, P. R. Modern Algorithms for Matching in Observational Studies. *Annual Review of Statistics and Its Application* **7** (2019).
34  Upham, S. P. & Small, H. Emerging research fronts in science and technology: patterns of new knowledge development. *Scientometrics* **83**, 15-38, doi:10.1007/s11192-009-0051-9 (2010).
35  Ataman, L. M., Ma, Y., Duncan, F. E., Uzzi, B. & Woodruff, T. K. Quantifying the growth of oncofertility. *Biology of reproduction* **99**, 263-265 (2018).
36  Bettencourt, L., Kaiser, D., Kaur, J., Castillo-Chavez, C. & Wojick, D. Population modeling of the emergence and development of scientific fields. *Scientometrics* **75**, 495-518 (2008).

37  Mane, K. K. & Börner, K. Mapping topics and topic bursts in PNAS. *Proceedings of the National Academy of Sciences* **101**, 5287-5290 (2004).
38  Small, H. Tracking and predicting growth areas in science. *Scientometrics* **68**, 595-610, doi:10.1007/s11192-006-0132-y (2006).
39  Abadie, A. Semiparametric difference-in-differences estimators. *The Review of Economic Studies* **72**, 1-19 (2005).
40  Abadie, A., Diamond, A. & Hainmueller, J. Synthetic control methods for comparative case studies: Estimating the effect of California's tobacco control program. *Journal of the American statistical Association* **105**, 493-505 (2010).
41  Allison, P. D. & Long, J. S. Departmental effects on scientific productivity. *American sociological review*, 469-478 (1990).
42  Borner, K. *Atlas of Knowledge*.  (MIT Press, 2014).
43  Petersen, A. M., Wang, F. & Stanley, H. E. Methods for measuring the citations and productivity of scientists across time and discipline. *Physical Review E* **81**, 036114 (2010).
44  Zuckerman, H. Nobel laureates in science: Patterns of productivity, collaboration, and authorship. *American Sociological Review*, 391-403 (1967).
45  Millar, M. M. Interdisciplinary research and the early career: The effect of interdisciplinary dissertation research on career placement and publication productivity of doctoral graduates in the sciences. *Research Policy* **42**, 1152-1164 (2013).
46  Azoulay, P., Zivin, J. S. & Wang, J. Superstar Extinction. *Quarterly Journal of Economics* **125**, 549-589 (2010).
47  Myers, K. The elasticity of science. *American Economic Journal: Applied Economics* (2020).
48  Wang, Y., Jones, B. F. & Wang, D. Early-career setback and future career impact. *Nature Communications* **10**, 4331, doi:10.1038/s41467-019-12189-3 (2019).
49  Park, H., Lee, J. J. & Kim, B.-C. Project selection in NIH: A natural experiment from ARRA. *Research Policy* **44**, 1145-1159 (2015).
50  Abadie, A., Diamond, A. & Hainmueller, J. Comparative politics and the synthetic control method. *American Journal of Political Science* **59**, 495-510 (2015).
51  Kreif, N. *et al.* Examination of the synthetic control method for evaluating health policies with multiple treated units. *Health economics* **25**, 1514-1528 (2016).
52  Merali, Z. Science prizes: The new Nobels. *Nature News* **498**, 152 (2013).
53  Fortunato, S. Prizes: Growing time lag threatens Nobels. *Nature* **508**, 186, doi:10.1038/508186a (2014).
54  Raftery, A. E. Bayesian model selection in social research. *Sociological methodology*, 111-163 (1995).
55  Kuhn, T. S. The essential tension: Tradition and innovation in scientific research. *Scientific creativity: Its recognition and development. New York: Wiley*, 341-354 (1963).
56  Wuchty, S., Jones, B. F. & Uzzi, B. The increasing dominance of teams in production of knowledge. *Science* **316**, 1036-1039 (2007).
57  Jones, B. F., Wuchty, S. & Uzzi, B. Multi-University Research Teams: Shifting Impact, Geography, and Stratification in Science. *Science* **322**, 1259-1262, doi:DOI 10.1126/science.1158357 (2008).

58	Uzzi, B., Mukherjee, S., Stringer, M. & Jones, B. Atypical combinations and scientific impact. *Science* **342**, 468-472 (2013).
59	Mukherjee, S., Romero, D. M., Jones, B. & Uzzi, B. The nearly universal link between the age of past knowledge and tomorrows breakthroughs in science and technology: The hotspot. *Science Advances* **3**, 2017, doi:e1601315 (2017).
60	Guimerà, R., Uzzi, B., Spiro, J. & Amaral, L. A. N. Team Assembly Mechanisms Determine Collaboration Network Structure and Team Performance. *Science* **308**, 697-702 (2005).
61	Jin, C., Song, C., Bjelland, J., Canright, G. & Wang, D. Emergence of scaling in complex substitutive systems. *Nature human behaviour* **3**, 837-846 (2019).
62	Zuckerman, H. The scientific elite: Nobel laureates' mutual influences. *Genius and eminence*, 157-169 (1992).
63	Oliveira, D. F. M., Ma, Y., Woodruff, T. K. & Uzzi, B. Comparison of National Institutes of Health Grant Amounts to First-Time Male and Female Principal Investigators. *JAMA* **321**, 898-900 (2019).
64	Lyall, C., Bruce, A., Marsden, W. & Meagher, L. The role of funding agencies in creating interdisciplinary knowledge. *Science and Public Policy* **40**, 62-71 (2013).
65	Börner, K. *et al.* A Multi-Level Systems Perspective for the Science of Team Science. *Science Translational Medicine* **2**, 49cm24, doi:10.1126/scitranslmed.3001399 (2010).
66	Gibney, E. What the Nobels are--and aren't--doing to encourage diversity. *Nature* **562**, 19-20 (2018).
67	Wikipedia. *International Congress of Mathematicians*, <https://en.wikipedia.org/wiki/International_Congress_of_Mathematicians> (2021).
68	Bertrand, M., Duflo, E. & Mullainathan, S. How much should we trust differences-in-differences estimates? *The Quarterly journal of economics* **119**, 249-275 (2004).
69	Stuart, E. A. Matching methods for causal inference: A review and a look forward. *Statistical science: a review journal of the Institute of Mathematical Statistics* **25**, 1 (2010).
70	Lu, S. F., Zhe Jin, G., Uzzi, B. & Jones, B. The Retraction Penalty: Evidence from the Web of Science. *Nature Scientific Reports* **3** (2013).

# Scientific Prizes and the Extraordinary Growth of Scientific Topics

**Authors:** Ching Jin[1,2], Yifang Ma[1,3], Brian Uzzi*[1,2]

**Affiliations:**

[1]Northwestern Institute on Complex Systems (NICO), Northwestern University, Evanston IL, 60208, USA.
[2]Kellogg School of Management, Northwestern University, Evanston IL, 60208, USA.
[3]Department of Statistics and Data Science, Southern University of Science and Technology, Shenzhen, Guangdong 518055, China
*Corresponding Author: uzzi@northwestern.edu

# Supplementary Information

## 1. Data Description:

**Scientific Prizes and Scientific Topics.** We collected a comprehensive dataset that combines 458 recognized scientific prizes with 12,041 scientific topics through prizewinning events occurred between 1970 and 2007. From these 12,041 prizewinning topics, more than 95% could be matched successfully with five non-prizewinning topics, results in our main dataset of 11,539 prizewinning topics and their related 405 prizes (for details see Sec. 4). The prizes in our sample include celebrated awards like the Wolf Prize and Turing Prize as well as hundreds of others recognized on Wikipedia's "scientific prizes" page. To validate the Wikipedia data, we manually cross-checked it with prize-related data on the web and in print media. To avoid misclassification of scientific prizes, a prize was considered a science prize if it had at least ten scientist-prizewinners following the Ma et al methodology.[1]

To link prizes awarded to scientists to scientific topics, we used Microsoft Academic Graph (MAG) data. MAG uses state of art NLP algorithms to classify 172,037,947 papers from 209,404,413 scientists to scientific topics before 2018[2,3]. The MAG associates papers with 228,251 topics (a.k.a "fields of study"), that are nested within 293 domains (e.g., Quantum Mechanics, Algebra, etc.), which are nested with 19 disciplines (e.g., Physics, Mathematics, etc.). Because the algorithm that assigns a paper to topics based on a paper's complete text and in relation to other texts, a paper's assigned topics are not synonymous with author-defined "keywords." For example, the paper, "The human disease network[4]," has the keywords biological networks, complex networks, human genetics, systems biology, and diseasome and classified firstly in the topics: Human Ineractome and then network medicine.

The above process generated a list of research topics for each publication of every prizewinner. To avoid misclassification errors of topic-labeling in MAG and to make sure that the listed topics are a meaningful part of a scientist's reputation, we considered a topic to be associated with a scientist prizewinner only if the scientist published at least $L = 10$ papers on the topic. We validated this criterion using Wikipedia's "*known for*" dataset, which crowdsources scientist opinions about the topics that other scientists are known for. The *known for* data has 3,427 "known-for" topics for our prizewinners. The average number of publications of a scientist's known-for topics is 10.0252, corroborating *L=10*. After further matching with non-prizewinning topics, 405 prizes are identified to be linked to 11,539 prizewinning topics through 2,900 prize conferrals. This dataset has been used in our main analyses. We also use the full 458-prize dataset in different robustness tests when selecting $L = 5, 15, and\ 20$, finding consistent results (Supplementary Fig. S1).

**NIH Grants:** We collected NIH grant data from 1985 to 2015 that estimated the amount of funding a topic received before and after the prize year. Grants were linked to different research topics through the NIH publication list associated with each grant. By counting the number of NIH grant links, we obtained an estimated "NIH grant mention" for each topic. Because the total number of NIH grants change year-by-year ($G(t)$) and, by definition, the NIH mention will change with this number, we adjusted the NIH mention by normalizing it with the total number of NIH grants ($G(t)$)

$$Adjusted\ NIH\ mention\ (t) = \frac{NIH\ mention\ (t)}{G(t)} * <G(t)>$$

where $G(t)$ is the total number of NIH grants for year $t$, and $<G(t)>$ calculates the average total number of NIH grants for different years. We find the adjusted NIH mention for the prizewinning topic is flat before and after the prizewinning event, indicating the funding has no measurable influence on the post-prize growth.

## 2. Quantifying Growth Patterns of Research Topics

To study whether a growth pattern before and after a topic is associated with a topic's prizewinning event, we find five topics in the same discipline that had growth patterns that were statistically indistinguishable from the prizewinning topics for ten years before the prize year.

To quantify the comparative growth of the prizewinning relative to the peer topics, we defined $\Delta_t$ over time $t$:

$$\Delta_t = \log(Y_t) - \log(\widetilde{Y}_t).$$

which measures the differences in the logarithm of the quantities and provides an appropriate method for establishing percentages changes in a quantity (prizewinning topic's growth, $Y_t$) relative to a baseline (geometric mean of the peer topics growth, $\widetilde{Y}_t$) since quantities such as citations may have fat-tail distributions[5,6]. As it is well known, the absolute differences perform better for quantities with a narrow Gaussian like distribution, which can be achieved by taking the logarithm of the six quantities, leading to our logarithm difference measure.

**Robustness Check:** To make sure our results are not methodology-dependent, we also test our results with standard ratio difference measure[7,8] (Supplementary Fig.S2 a-f), which show growth patterns similar to the ones we present in the main text. A further placebo test also corroborates our main findings (Supplementary Fig.S2 g-l). Specifically, for each of the prizewinning topic, we selected one non-prizewinning topic from its peer topic candidate pool as a "pretend winning topic", repeating the analysis in Supplementary Fig.2 a-f, finding that there is no difference in the expected growth for pretend topic before and after the prizewinning event, supporting our main conclusions.

## 4. Dynamic Optimal Matching Method

To select the peer non-prizewinning topics, we use a Dynamic Optimal Matching method, which applies the Optimal Matching Method[9-11] to a time-series data to simultaneously maximize the closeness and balance characteristics of accurate matching[10-13]. We focus on prizewinning between 1970 and 2007, resulting in a sample of 12,041 prizewinning and peer non-prizewinning topics. 11,539 topics (> 95%) could be statistically matched to peer topics using DOM. If a topic has multiple prizes over its lifetime, we match for the first prizewinning.

First, we select a peer topic candidate pool[9]. To achieve this, for each prizewinning topic $i$, we selected up to 40 close-distance topics in terms of a distance measure ($\theta_{i,j}$) from the same discipline, generating a peer candidate pool. For 95% of the prizewinning topics, a proper peer candidate pool was identified (11,539/12,041= 95.8%). To achieve matching, we defined a distance measure $\theta_{i,j}$ to quantify the closeness between the prizewinning topic $i$ and a non-prizewinning topic $j$[7,8]:

$$\theta_{i,j} = \frac{\sum_{n=1}^{N} \sum_{t=t^*-t_0}^{t^*} \left(\log Y_{i,n}(t) - \log Y_{j,n}(t)\right)^2}{N * (t_0 + 1)}$$

, where $Y_{i,n}$ indicates the quantity for the topic $i$ in terms of one of the $N=6$ matched categories (i.e. Productivity, Citations, Lead Scientist impact, #incumbents, #Entrants and #Disciplinary Stars). $t$ measures number of years prior to the prizewinning year. $t^*$ represents the prizewinning year for topic $i$, and $t_0 = 10$, indicates we traced the growth pattern for topics in an 11-year duration, which includes 10 years prior to the prizewinning year.

Second, to ensure the balance between the peer and prizewinning topics for the entire system, we select 5 matching topics from the candidate pool to be the topic's peer group. In this process, we (1) minimize the distances between the peer and prizewinning topics in terms of $\theta_{i,j}$, and (2) make sure the distribution of the peer and prizewinning topics are acceptably and simultaneously close for all 66 covariates.

Specifically, we make sure the differences between the prizewinning and peer topic groups are small enough for each matching category n, and for any time $t$ before the prizewinning event ($-10 \leq t \leq 0$), where the differences between the prizewinning topic $i$ and its expected growth at time $t$ and category $n$ are quantified by $\Delta_{i,n}(t) = \left(\log Y_{i,n}(t) - \log \tilde{Y}_{i,n}(t)\right)$. The expected growth is obtained by averaging the trajectory of the matched topics. This problem is a classical optimization problem, which could be solved with typical Mixed Integer Programming (MIP) methods[9,13]. We found the best-optimized matching possible where (1) the distance between the prizewinning topics and the peer topics is minimized; at the same time, (2) the difference between the peer and prizewinning groups is not statistically significant for any $t \in [-10,0]$ and $n \in [1,6]$. Mathematically, we have:

$$\left|\frac{\sum_{i=1}^{M} \Delta_{i,n}(t)}{M} - 0\right| < 1.96 * SE(\Delta_{i,n}(t)). \ for \ \forall \ 1 \leq n \leq 6, -10 \leq t \leq 0$$

Here $SE(\Delta_{i,n}(t))$ measures the standard error of the $\Delta_{i,n}(t)$ for the prizewinning topics at time $t$ and in category $n$, and $M$ captures number of prizewinning topics. To prevent bias by topics with a large $\Delta_{i,n}(t)$ in the MIP process, we also monitored the topic-by-topic growth of each individual topic. Specifically, for any $t$ ($-10 \leq t \leq 0$), we compare the growth pattern of each prizewinning topic and all of its peer topics (11,539*5=57,695 pairs). For each of the 66 covariances, we ensure each prizewinning topic had equal probability to grow faster or slower than any of its peer topics. Supplementary Fig. S4 shows the binomial tests validating our method, demonstrating that all p-value is larger than 0.2. This method not only guarantees closeness between the peer and prizewinning topics but also unsure good balancing *between* and *within* topic groups.

To further validate our DOM method, we also run an additional "placebo" tests that pretend that each peer topic is a prizewinning topic with the purpose of testing whether the peer topics also showed abnormal growth following the prize year of the prizewinning topic (Supplementary Fig. S5). Specifically, for each of the prizewinning topic, we selected a non-prizewinning topic from its matching candidates as a "pretend winning topic", matched it with 5 peer topics, repeating the whole analysis in Fig. 2, finding that peer topics have no coincidental extraordinary growth (all *p*-values>0.05), reinforcing our main finding prizewinning is associated with and a topic's onset of a sustained period of extraordinary.

To account for possible correlations among different variables, we also repeat the analysis by adopting *Mahalanobis distance* to quantify the closeness of the topics. Specifically, we present the pre-prize growth patterns of a topic as a 66-element (6 categories * 11 years) vector $\vec{y_i}$, quantifying the logarithm growth of the topic $i$ in the 11-years period before the prizewinning event in terms of the six measures. We can calculate the Mahalanobis distance between the prizewinning topic $i$ and a non-prizewinning topic $j$:

$$\theta'_{i,j} = \sqrt{(\vec{y_i} - \vec{y_j})^T S^{-1} (\vec{y_i} - \vec{y_j})},$$

where $S$ is the covariance matrix. By adopting this distance measure in our matching procedure, we repeat the main analysis, finding consistent results (Supplementary Fig. S6).

## 5. Paradigmatic Diversification

Prizewinning topics become paradigmatically diverse than matched peer topics. Paradigmatic diversification refers to the heterogeneity of concepts scientists use to study a topic[14,15]. To measure paradigmatic diversification, we created a master list of all the topics new entrants in prizewinning or peer topics had published on before becoming a new entrant. Topics in the master list were defined as different from one another if topics were associated with different disciplines (N=19 disciplines). The topic diversity of the master list was then measured using Shannon Entropy as: $S = -\sum_j p_j \log_2 p_j$, where $j$ represents a discipline, and $p_j$ measures the probability that a topic in the list belongs to discipline $j$. We observe that paradigmatic diversity's distribution at $\Delta_{10} \geq 0$ for prizewinning topics is significantly more diverse than peer topics (K-S test, $\Delta_{10} \geq 0$ group, $p < 0.0001$). Supplementary Fig. S7 shows that when $\Delta_{10}$ equals zero (i.e., no growth differences), prizewinning and peer topics have no significant difference in paradigmatic diversification. However, as $\Delta_{10}$ grows, relative paradigmatic diversification increases significantly (slope $= 0.109$, $p < 0.0001$). For example, when $\Delta_{10}$ equals to 1.5, paradigmatic diversity is 11.6% greater for prizewinning topics than it is for peer topics.

## 6. Prize Characteristics Predict Magnitude of Extraordinary Growth

To examine the results' sensitivity to confounds, we regressed the $\Delta_{10}$ of our six growth variables on money, discipline-specific, and recency along with control variables. Control variables include lagged values of each growth trend at times *t*-1, *t*-2, and *t*-3 years to account for autoregressive effects of $\Delta_{10}$. A disciplinary fixed effect variable controls for stable disciplinary differences such as prestige, theoretical vs bench science disciplines, and level of paradigm development. To control for yearly numbers of scientists and publications, we added a calendar year fixed effect. To account for differences in a prize's visibility, we added control variables for total Wikipedia pageviews of the prize (measured as of 2017), a binary variable for whether the prizewinner is among the top 5% of cited authors on that topic, a binary variable for whether there are multiple prize recipients for the topic in the same year, number of prize conferrals up to the prize year, and the age of the prize up to the prize year. To better interpret the results, we show both results of the original variables (Supplementary Tab. S2) and the variables which have been standardized (Supplementary Tab. S3).

The results are further validated by adopting the *BIC* statistics, where we find signal strengths adding a large explanation to the extraordinary growth of prizewinning topics (Supplementary Tab. S4-S9). 10-fold cross-validation indicated that there was no overfitting of the regression model (Supplementary Tab. S10). Because the amount of prize money can range widely, and its non-normal cannot be linearized through mathematical transformations, we also do a further regression test by creating a three-category money variable defined as (a) no money, (b) money below the median, and (c) money above the median (Supplementary Tab. S11), which confirmed the simpler binary variable result which we report. A list and description of variables details please see Supplementary Tab. S12.

# SI Figures

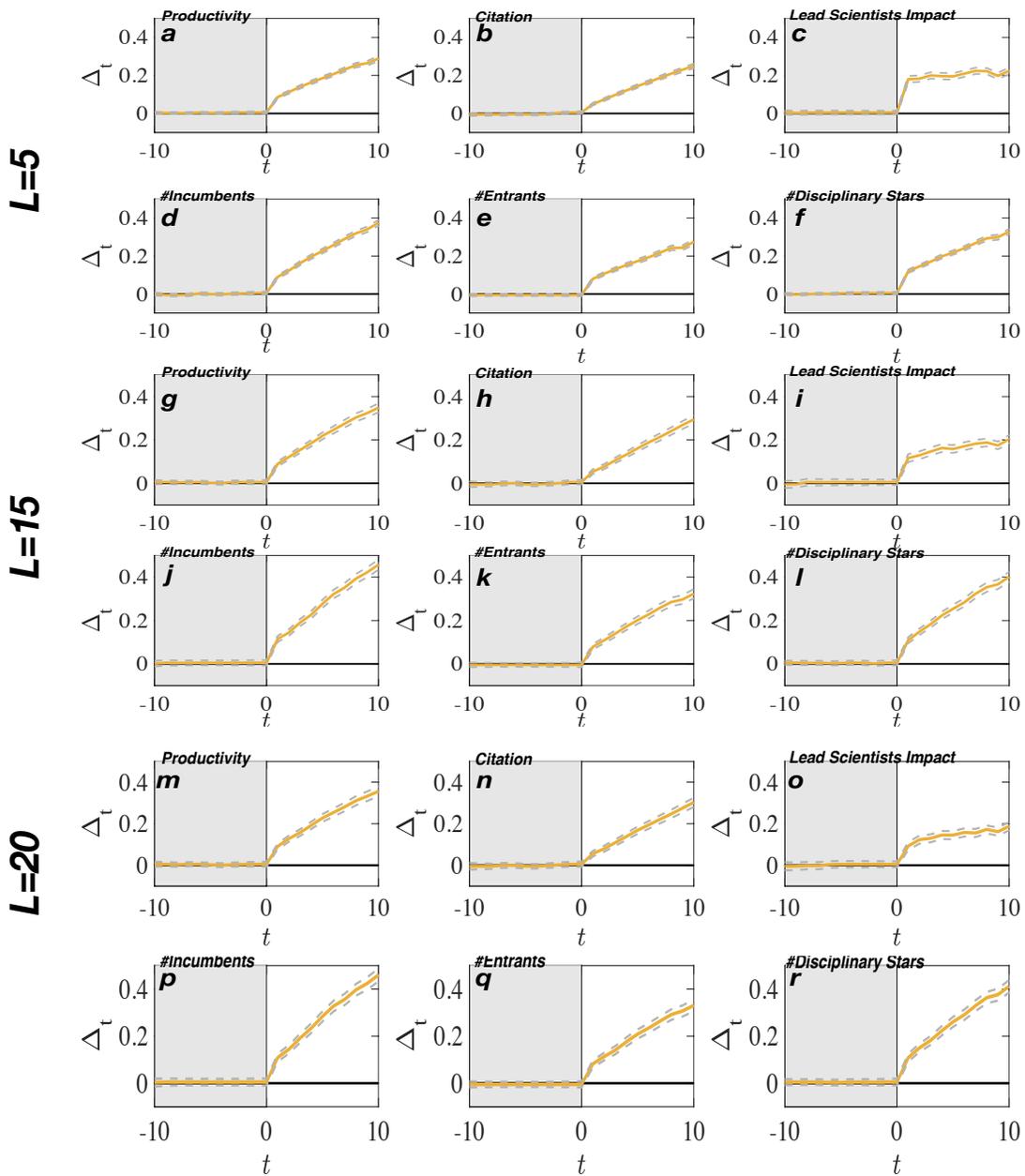

**Supplementary Fig. S1. Main analysis with "known for" topics measured as >=5, >=15 or >=20 papers on a topic as opposed to >=10 papers. (a-r)** Here we use an alternative criterion of the main topics for prizewinners by selecting **(a-f)** $L = 5$, **(g-l)** $L = 15$, and **(m-r)** $L = 20$. The findings are consistent with the main results. The dashed lines indicate the 95% confidence intervals.

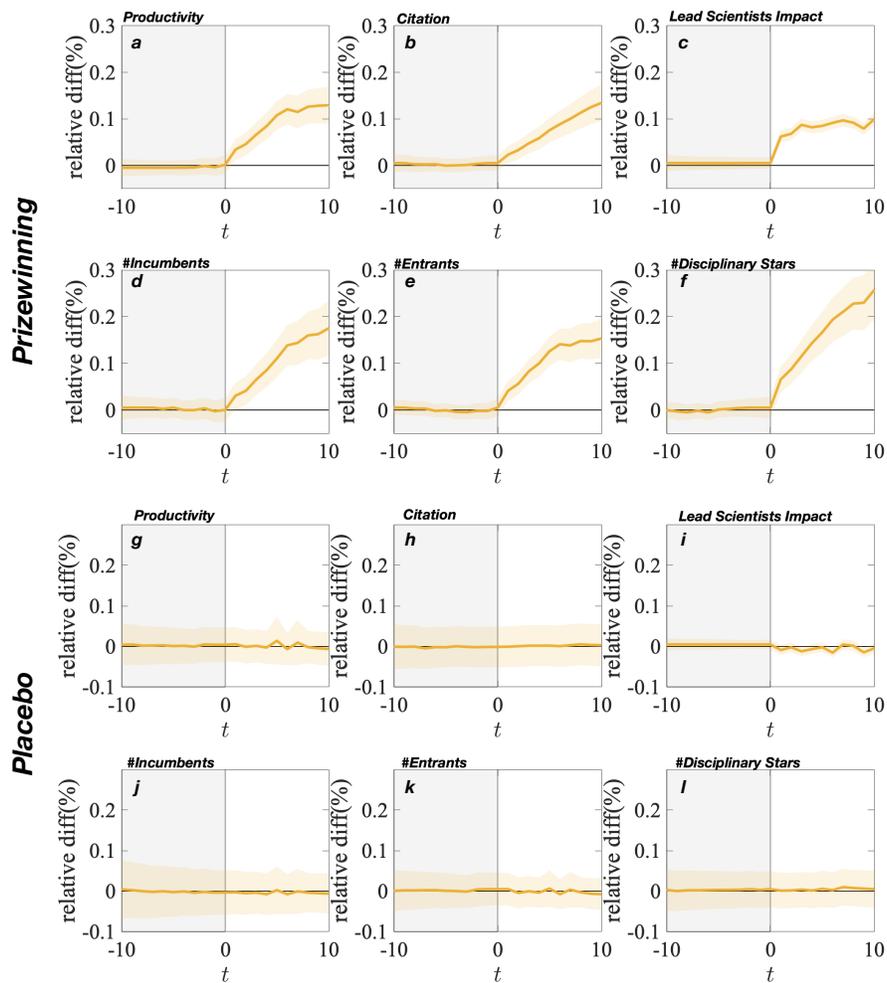

**Supplementary Fig. S2. Alternative measures of extraordinary growth. (a-f)** We repeat the main results with the alternative definition of difference. Here we calculate the ratio (relative) difference between the prizewinning topics and the expected growth, finding the main findings remain the same. **(g-l)** Placebo test for the analysis for results in a-e, where the growth pattern of the peer topic is flat before and after the prize. Error band indicates the 95% CI.

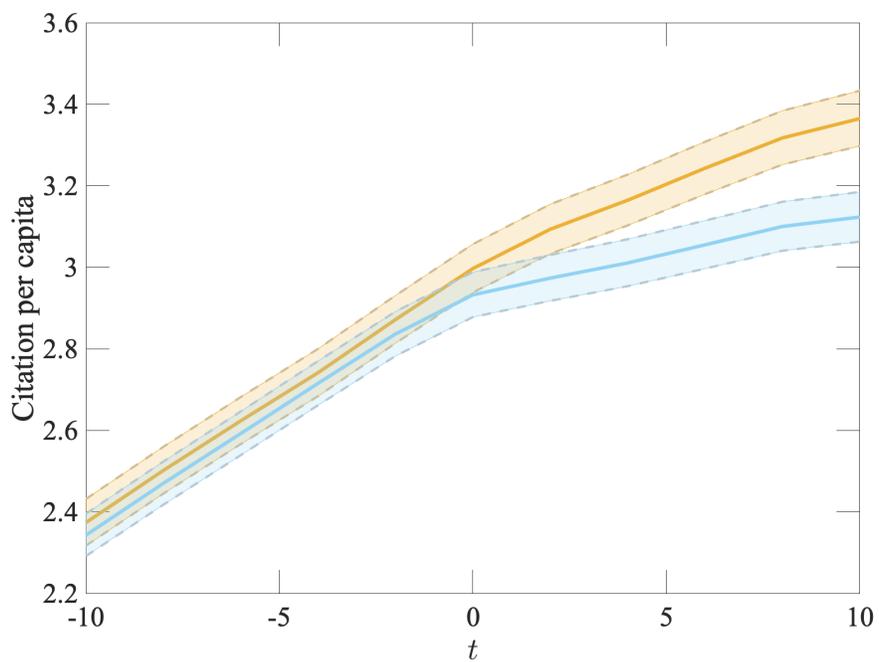

**Supplementary Fig. S3. Citation per capita.** Citations per capita increase after the prize, indicating the prize is also an important signal for research quality. Yellow curve corresponds to the prizewinning topics, and the blue curve represents the matched topics. Error band measures 95% CI.

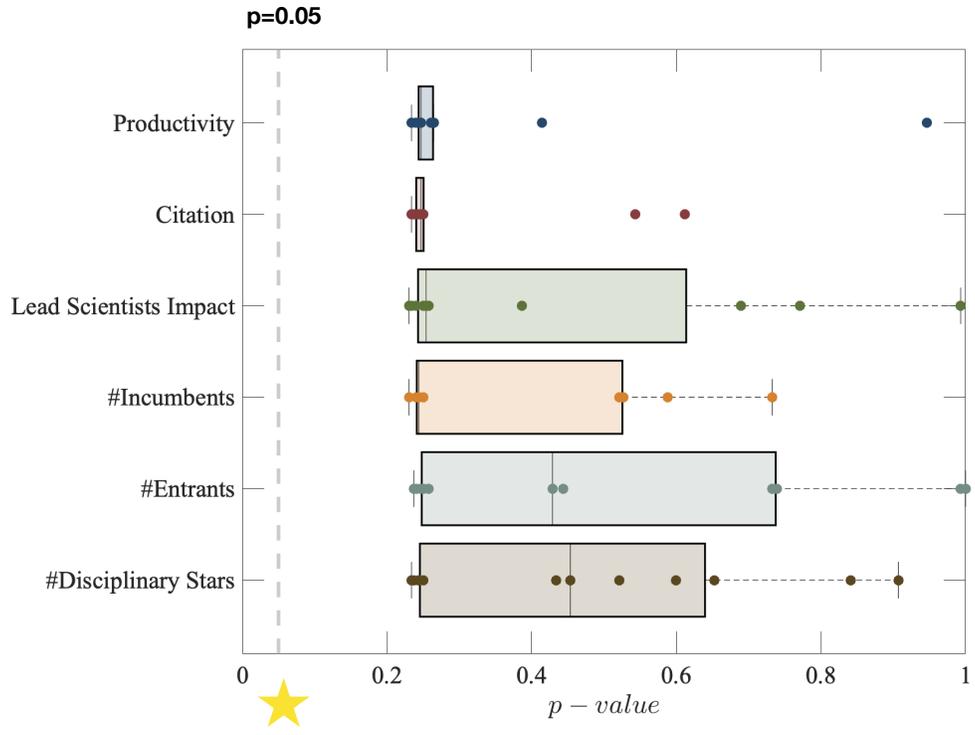

**Supplementary Fig. S4. Prizewinning and Matched Non-prizewinning Peer Topics Have Statistically Indistinguishable Historical Growth Patterns Prior to the Prize Year.** Based on six measures of a topic's growth (productivity, citation, lead scientist impact, #incumbents, #entrants and #disciplinary stars), we observed that for 11 consecutive years before the year the prizewinning topic is awarded its prize, the prizewinning topic and its peer topic have statistically indistinguishable growth patterns. Box plots show median and 90[th] and 10[th] percentiles of the p-value of the 11 two-tailed binomial tests, one for each year prior to the prize year. N=57,695 (11,539*5) topic pairs are observed for individual test. All tests of all measures (11*6==66 measures, dots) have a p-value > 0.05 (dashed line). The center line of the box plot is the median of the normalized grants, box limits correspond to the data's first and third quartiles, notches represent 95% CI,

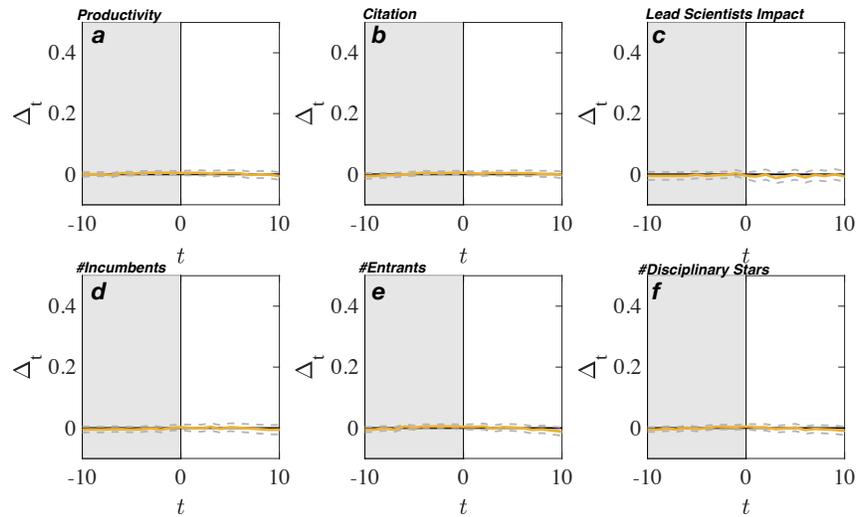

**Supplementary Fig. S5. Placebo Tests.** For each prizewinning topic, we select a non-prizewinning topic from the matching candidate pool as a "fake prizewinning topic". We repeat the DOM matching process for this faked topic, and search for the five peers for it. We perform the same analysis as we did for the prizewinning topic, finding that for these faked topics, there is no significant difference before and after the prizewinning event, validating our DOM method.

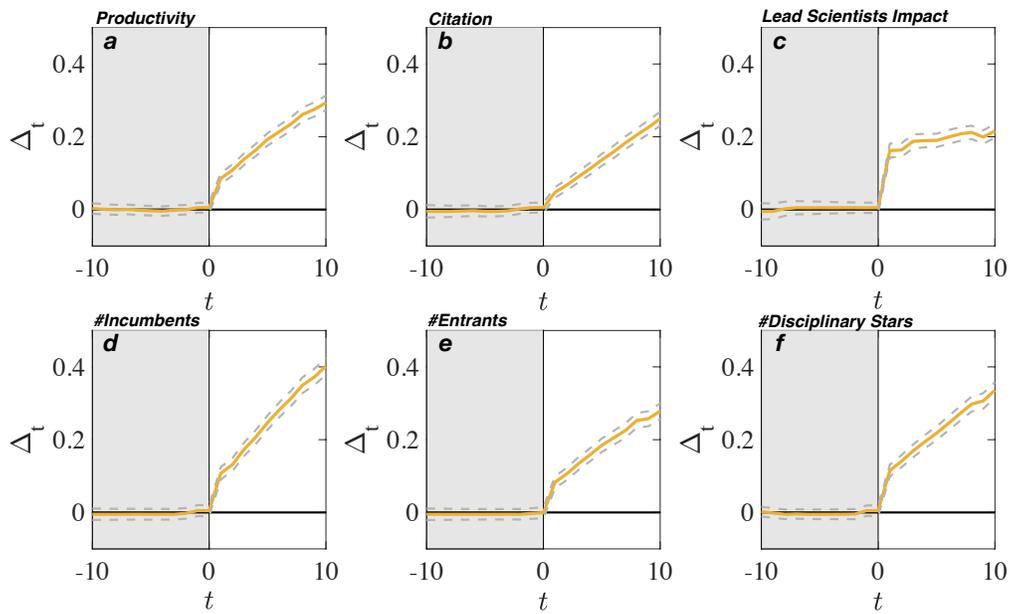

**Supplementary Fig. S6. Using Mahalanobis distance in the matching procedure. (a-f)** We repeat the main results with the alternative measure (Mahalanobis distance) to quantify the closeness of topics. We rematch each prizewinning topic with this new distance measure, repeating the main analysis, finding consistent growth patterns.

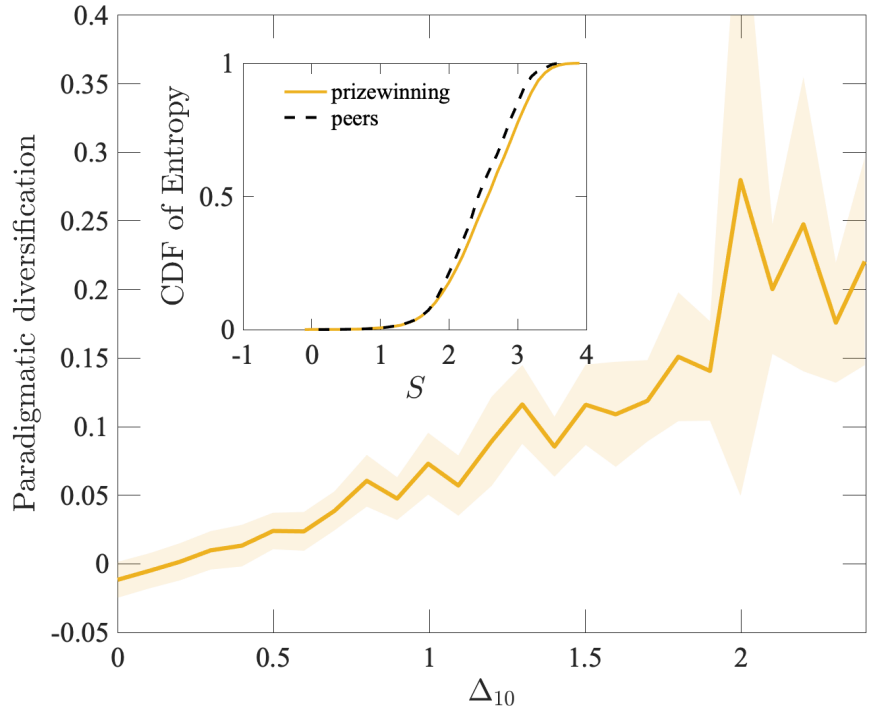

**Supplementary Fig. S7. Prizes and Paradigmatic Diversification.** The plot shows the percentage increase in paradigmatic diversification for prizewinning topics relative to peer topics. The inset shows the cumulative distribution of the paradigmatic diversity of the prizewinning topics and their peer groups significantly differ (K-S tst, two tailed test, $p = 3.1 * 10^{-25}$). The relationship between extraordinary growth and paradigmatic diversity shows that as $\Delta_{10}$ increases, relative paradigmatic diversity increases significantly (slope=0.109, OLS Regression, $p = 8.7 * 10^{-218}$). For example, when $\Delta_{10}$ equal to 1.5, a prizewinning topic is estimated to be 11.6% more diverse paradigmatical than its peer topics. Error band corresponds to 95% CI.

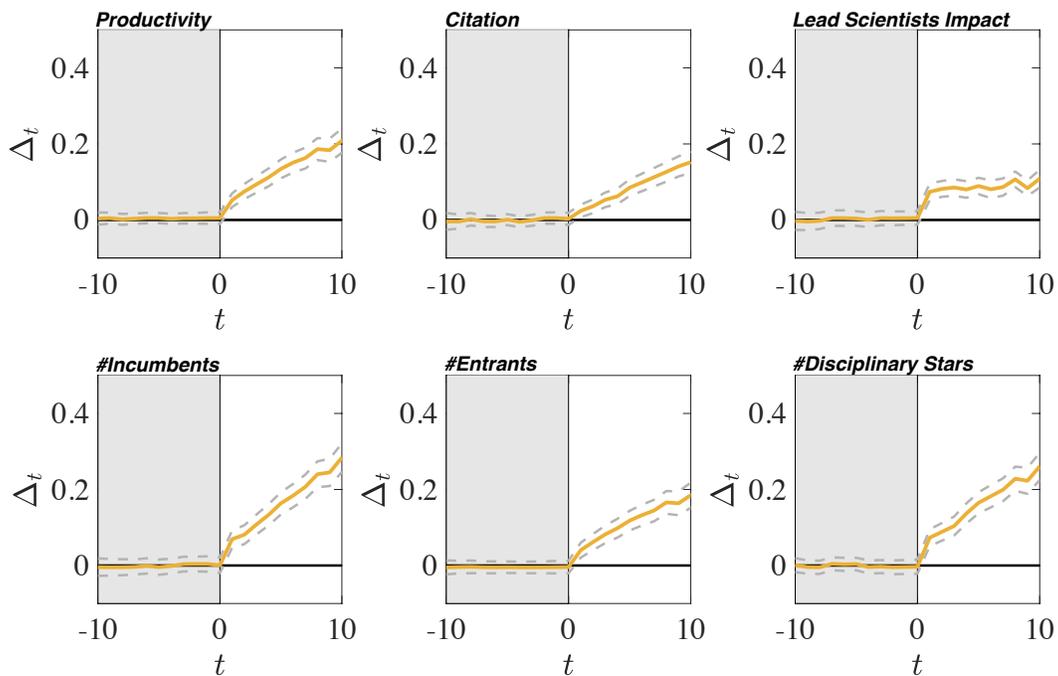

**Supplementary Fig. S8. NIH subsample analysis (both prizewinning and peer topics are NIH funded).** This figure reports a new analysis where both the prizewinning and peer topics are NIH funded topics. Of the 2,853 NIH funded prizewinning topics, 2,569 prizewinning topics were matched with five non-prizewinning, NIH funded peer topics using DOM. This procedure ensures that 100% of the peer topics are also NIH funded topics and the subsample is balanced and meets the parallel trend requirement. By repeating this matching process in this NIH funded topic pool, we find prizewinning topics grow relatively larger than peer topics after the prize year, in consistent with the main results reported in the main manuscript.

|  | Differences Between Prizewinning and Peer Topics' Big Five Growth Measures for 10 Years Before and After Prize | |
|---|---|---|
|  | 10 years Before the Prize | 10 years After the Prize |
| Annual Growth Measure | Binomial test | Binomial test |
| 1. productivity | 0.5024 (ns) | 0.6159*** |
| 2. citation | 0.5025 (ns) | 0.5976*** |
| 3. lead scientist impact | 0.5008 (ns) | 0.5821*** |
| 4. #incumbents | 0.5024 (ns) | 0.6238*** |
| 5. #entrants | 0.5007 (ns) | 0.5987*** |
| 6. #disciplinary stars | 0.5025 (ns) | 0.6177*** |
| All tests 2-tailed. Binomial tests *** p<0.001 | | |

**Supplementary Table S1. Binomial tests for extraordinary growths after prizewinning events.** For each of our 11,539 prizewinning topics and their peer topics, we calculate the fraction of pairs where the prizewinning topic grows faster than the peer topic, we find around 60% of the prizewinning topics are likelier to growth faster than their specific peer topics (binomial test, all *p* values<0.001), indicating that the link between prizewinning and extraordinary growth generalizes on the topic-by-topic level. As a comparison, the same analysis has been performed for the data 10 years before the prize, we find the fraction of topics with a positive growth is no significant different with 0.5 (two-tailed binomial test, all p values> 0.2, specific p-value see Fig. S4), indicating the difference between the prizewinning topics and the peer topics are well balanced before the prizewinning year, further validating our DOM method.

|  | Productivity | Citations | Impact of Topic's Lead Scientists | #Incumbents | #Entrants | #Disciplinay Stars working on the topic |
|---|---|---|---|---|---|---|
| **Recency** | 0.0139*** | 0.0134*** | 0.00905*** | 0.0167*** | 0.0141*** | 0.0146*** |
|  | (0.000978) | (0.000955) | (0.000936) | (0.00116) | (0.00103) | (0.00109) |
| **Money (Yes/No)** | 0.0393** | 0.0329* | -0.0210 | 0.0413* | 0.0391* | 0.0401* |
|  | (0.0146) | (0.0131) | (0.0162) | (0.0177) | (0.0157) | (0.0166) |
| **Discipline/General** | 0.0705*** | 0.103*** | 0.0520** | 0.0869*** | 0.0876*** | 0.127*** |
|  | (0.0165) | (0.0151) | (0.0195) | (0.0202) | (0.0177) | (0.0189) |
| **Prizewinner top** | -0.0239 | -0.0151 | -0.0126 | -0.0189 | -0.0283 | -0.0234 |
|  | (0.0158) | (0.0141) | (0.0167) | (0.0192) | (0.0168) | (0.0176) |
| **Prize Age** | 0.0000716 | 0.0000989 | 0.00000621 | 0.000182 | 0.0000124 | 0.0000498 |
|  | (0.0000830) | (0.0000782) | (0.000122) | (0.0000935) | (0.0000933) | (0.000114) |
| **Conferral times** | 0.000286 | 0.000672*** | 0.000140 | 0.000260 | 0.000275 | 0.000331 |
|  | (0.000207) | (0.000181) | (0.000203) | (0.000245) | (0.000221) | (0.000235) |
| **Pageviews** | 0.00000251 | 0.00000176 | -0.000000124 | 0.00000333 | 0.00000290* | 0.00000252 |
|  | (0.00000138) | (0.00000119) | (0.00000159) | (0.00000180) | (0.00000147) | (0.00000167) |
| **Discipline** | Yes | Yes | Yes | Yes | Yes | Yes |
| **Prizewinning year** | Yes | Yes | Yes | Yes | Yes | Yes |
| **Multiple recipients** | Yes | Yes | Yes | Yes | Yes | Yes |
| **Lagged variable** | Yes | Yes | Yes | Yes | Yes | Yes |
| **const** | 0.662*** | 0.482*** | 0.542*** | 0.827*** | 0.703*** | 0.700*** |
|  | (0.0880) | (0.0774) | (0.123) | (0.101) | (0.0991) | (0.101) |
| **R-square** | 0.318 | 0.374 | 0.120 | 0.267 | 0.295 | 0.299 |

* $p<0.05$, ** $p<0.01$, *** $p<0.001$

**Supplementary Table S2: Regression analysis with robust standard errors of the relations between Prize Characteristics and Magnitude of Extraordinary Growth. OLS regression are used in these analyses.**

|  | Productivity | Citations | Impact of Topic's Lead Scientists | #Incumbents | #Entrants | #Disciplinay Stars working on the topic |
| --- | --- | --- | --- | --- | --- | --- |
| **Recency** | 0.143*** | 0.146*** | 0.092*** | 0.146*** | 0.138*** | 0.134*** |
|  | (0.000978) | (0.000955) | (0.000936) | (0.00116) | (0.00103) | (0.00109) |
| **Money (Yes/No)** | 0.021** | 0.019* | -0.011 | 0.019* | 0.020* | 0.020* |
|  | (0.0146) | (0.0131) | (0.0162) | (0.0177) | (0.0157) | (0.0166) |
| **Discipline/General** | 0.034*** | 0.053*** | 0.025** | 0.036*** | 0.040*** | 0.055*** |
|  | (0.0165) | (0.0151) | (0.0195) | (0.0202) | (0.0177) | (0.0189) |
| **Prizewinner top** | -0.012 | -0.008 | -0.006 | -0.008 | -0.013 | -0.010 |
|  | (0.0158) | (0.0141) | (0.0167) | (0.0192) | (0.0168) | (0.0176) |
| **Prize Age** | 0.005 | 0.008 | 0.000 | 0.011 | 0.001 | 0.003 |
|  | (0.0000830) | (0.0000782) | (0.000122) | (0.0000935) | (0.0000933) | (0.000114) |
| **Conferral times** | 0.011 | 0.028*** | 0.005 | 0.009 | 0.010 | 0.012 |
|  | (0.000207) | (0.000181) | (0.000203) | (0.000245) | (0.000221) | (0.000235) |
| **Pageviews** | 0.015 | 0.011 | -0.001 | 0.017 | 0.016* | 0.013 |
|  | (0.00000138) | (0.00000119) | (0.00000159) | (0.00000180) | (0.00000147) | (0.00000167) |
| **Discipline** | Yes | Yes | Yes | Yes | Yes | Yes |
| **Prizewinning year** | Yes | Yes | Yes | Yes | Yes | Yes |
| **Multiple recipients** | Yes | Yes | Yes | Yes | Yes | Yes |
| **Lagged variable** | Yes | Yes | Yes | Yes | Yes | Yes |
| **R-square** | 0.318 | 0.374 | 0.120 | 0.267 | 0.295 | 0.299 |

* $p<0.05$, ** $p<0.01$, *** $p<0.001$

**Supplementary Table S3: Regression analysis with robust standard errors of the relations between Prize Characteristics and Magnitude of Extraordinary Growth (Beta Coefficients). OLS regression are used in these analyses.**

|  | Productivity | Productivity | Productivity | Productivity | Productivity |
|---|---|---|---|---|---|
| **Recency** | 0.0140*** |  |  |  | 0.0139*** |
|  | (0.000977) |  |  |  | (0.000978) |
| **Money (Yes/No)** |  | 0.0554*** |  | 0.0565*** | 0.0393** |
|  |  | (0.0146) |  | (0.0146) | (0.0146) |
| **Discipline/General** |  |  | 0.0640*** | 0.0654*** | 0.0705*** |
|  |  |  | (0.0167) | (0.0167) | (0.0165) |
| **Prizewinner top** | -0.0213 | -0.00344 | -0.00541 | -0.00577 | -0.0239 |
|  | (0.0158) | (0.0159) | (0.0160) | (0.0159) | (0.0158) |
| **Prize Age** | 0.0000181 | 0.0000742 | 0.0000522 | 0.0000964 | 0.0000716 |
|  | (0.0000824) | (0.0000797) | (0.0000796) | (0.0000798) | (0.0000830) |
| **Conferral times** | 0.0000797 | -0.0000550 | 0.000178 | 0.000154 | 0.000286 |
|  | (0.000198) | (0.000198) | (0.000207) | (0.000207) | (0.000207) |
| **Pageviews** | 0.00000229 | 0.00000211 | 0.00000330* | 0.00000269 | 0.00000251 |
|  | (0.00000138) | (0.00000141) | (0.00000140) | (0.00000141) | (0.00000138) |
| **Discipline** | Yes | Yes | Yes | Yes | Yes |
| **Prizewinning year** | Yes | Yes | Yes | Yes | Yes |
| **Multiple recipients** | Yes | Yes | Yes | Yes | Yes |
| **Lagged variable** | Yes | Yes | Yes | Yes | Yes |
| const | 0.752*** | 0.601*** | 0.546*** | 0.529*** | 0.662*** |
|  | (0.0848) | (0.0859) | (0.0887) | (0.0889) | (0.0880) |
| R-square | 0.316 | 0.299 | 0.299 | 0.300 | 0.318 |
| $\Delta BIC$ | -312.74 | -5.05 | -4.33 | -9.97 | -318.04 |

\* $p<0.05$, \*\* $p<0.01$, \*\*\* $p<0.001$

**Supplementary Table S4: Regression analysis with robust standard errors of the relations between Prize Characteristics and Magnitude of Extraordinary Growth (Productivity). OLS regression are used in these analyses.**

|  | Citations | Citations | Citations | Citations | Citations |
|---|---|---|---|---|---|
| **Recency** | 0.0134*** |  |  |  | 0.0134*** |
|  | (0.000956) |  |  |  | (0.000955) |
| **Money (Yes/No)** |  | 0.0478*** |  | 0.0495*** | 0.0329* |
|  |  | (0.0131) |  | (0.0131) | (0.0131) |
| **Discipline/General** |  |  | 0.0966*** | 0.0978*** | 0.103*** |
|  |  |  | (0.0152) | (0.0152) | (0.0151) |
| **Prizewinner top** | -0.0114 | 0.00583 | 0.00266 | 0.00235 | -0.0151 |
|  | (0.0142) | (0.0143) | (0.0143) | (0.0143) | (0.0141) |
| **Prize Age** | 0.0000398 | 0.0000896 | 0.0000840 | 0.000123 | 0.0000989 |
|  | (0.0000775) | (0.0000766) | (0.0000767) | (0.0000767) | (0.0000782) |
| **Conferral times** | 0.000358* | 0.000231 | 0.000565** | 0.000545** | 0.000672*** |
|  | (0.000174) | (0.000174) | (0.000184) | (0.000183) | (0.000181) |
| **Pageviews** | 0.00000118 | 0.00000107 | 0.00000247* | 0.00000194 | 0.00000176 |
|  | (0.00000118) | (0.00000121) | (0.00000120) | (0.00000121) | (0.00000119) |
| **Discipline** | Yes | Yes | Yes | Yes | Yes |
| **Prizewinning year** | Yes | Yes | Yes | Yes | Yes |
| **Multiple recipients** | Yes | Yes | Yes | Yes | Yes |
| **Lagged variable** | Yes | Yes | Yes | Yes | Yes |
| const | 0.0134*** | 0.0134*** | 0.0134*** | 0.0134*** | 0.0134*** |
|  | (0.000955) | (0.000955) | (0.000955) | (0.000955) | (0.000955) |
| R-square | 0.372 | 0.353 | 0.355 | 0.355 | 0.374 |
| $\Delta BIC$ | -355.67 | -3.85 | -29.10 | -33.92 | -387.55 |

\* $p<0.05$, \*\* $p<0.01$, \*\*\* $p<0.001$

**Supplementary Table S5: Regression analysis with robust standard errors of the relations between Prize Characteristics and Magnitude of Extraordinary Growth (Citations). OLS regression are used in these analyses.**

|  | Impact of Topic's Lead Scientists | Impact of Topic's Lead Scientists | Impact of Topic's Lead Scientists | Impact of Topic's Lead Scientists | Impact of Topic's Lead Scientists |
| --- | --- | --- | --- | --- | --- |
| **Recency** | 0.00894*** |  |  |  | 0.00905*** |
|  | (0.000927) |  |  |  | (0.000936) |
| **Money (Yes/No)** |  | -0.0106 |  | -0.00977 | -0.0210 |
|  |  | (0.0161) |  | (0.0161) | (0.0162) |
| **Discipline/General** |  |  | 0.0488* | 0.0486* | 0.0520** |
|  |  |  | (0.0196) | (0.0196) | (0.0195) |
| **Prizewinner top** | -0.0108 | 0.000953 | -0.000839 | -0.000777 | -0.0126 |
|  | (0.0166) | (0.0167) | (0.0168) | (0.0168) | (0.0167) |
| **Prize Age** | 0.00000564 | 0.00000588 | 0.0000300 | 0.0000224 | 0.00000621 |
|  | (0.000122) | (0.000118) | (0.000118) | (0.000118) | (0.000122) |
| **Conferral times** | -0.0000380 | -0.000102 | 0.0000498 | 0.0000538 | 0.000140 |
|  | (0.000192) | (0.000191) | (0.000202) | (0.000202) | (0.000203) |
| **Pageviews** | -0.000000823 | -0.000000436 | -0.000000106 | -2.02e-09 | -0.000000124 |
|  | (0.00000157) | (0.00000159) | (0.00000160) | (0.00000161) | (0.00000159) |
| **Discipline** | Yes | Yes | Yes | Yes | Yes |
| **Prizewinning year** | Yes | Yes | Yes | Yes | Yes |
| **Multiple recipients** | Yes | Yes | Yes | Yes | Yes |
| **Lagged variable** | Yes | Yes | Yes | Yes | Yes |
| const | 0.592*** | 0.509*** | 0.452*** | 0.455*** | 0.542*** |
|  | (0.120) | (0.120) | (0.122) | (0.122) | (0.123) |
| **R-square** | 0.120 | 0.112 | 0.113 | 0.113 | 0.120 |
| **ΔBIC** | -91.7 | 9.00 | 3.32 | 12.29 | -387.55 |

\* p<0.05, \*\* p<0.01, \*\*\* p<0.001

**Supplementary Table S6: Regression analysis with robust standard errors of the relations between Prize Characteristics and Magnitude of Extraordinary Growth (Impact of Topic's Lead Scientists). OLS regression are used in these analyses.**

|  | #Incumbents | #Incumbents | #Incumbents | #Incumbents | #Incumbents |
| --- | --- | --- | --- | --- | --- |
| **Recency** | 0.0167*** |  |  |  | 0.0167*** |
|  | (0.00116) |  |  |  | (0.00116) |
| **Money (Yes/No)** |  | 0.0606*** |  | 0.0620*** | 0.0413* |
|  |  | (0.0177) |  | (0.0177) | (0.0177) |
| **Discipline/General** |  |  | 0.0793*** | 0.0808*** | 0.0869*** |
|  |  |  | (0.0205) | (0.0205) | (0.0202) |
| **Prizewinner top** | -0.0157 | 0.00584 | 0.00336 | 0.00297 | -0.0189 |
|  | (0.0192) | (0.0193) | (0.0193) | (0.0193) | (0.0192) |
| **Prize Age** | 0.000121 | 0.000184* | 0.000163 | 0.000211* | 0.000182 |
|  | (0.0000927) | (0.0000921) | (0.0000919) | (0.0000922) | (0.0000935) |
| **Conferral times** | 0.00000193 | -0.000157 | 0.000127 | 0.000102 | 0.000260 |
|  | (0.000234) | (0.000235) | (0.000246) | (0.000246) | (0.000245) |
| **Pageviews** | 0.00000299 | 0.00000284 | 0.00000422* | 0.00000356 | 0.00000333 |
|  | (0.00000179) | (0.00000181) | (0.00000181) | (0.00000182) | (0.00000180) |
| **Discipline** | Yes | Yes | Yes | Yes | Yes |
| **Prizewinning year** | Yes | Yes | Yes | Yes | Yes |
| **Multiple recipients** | Yes | Yes | Yes | Yes | Yes |
| **Lagged variable** | Yes | Yes | Yes | Yes | Yes |
| **const** | 0.935*** | 0.756*** | 0.686*** | 0.667*** | 0.827*** |
|  | (0.0978) | (0.0991) | (0.102) | (0.102) | (0.101) |
| **R-square** | 0.266 | 0.248 | 0.248 | 0.249 | 0.267 |
| **ΔBIC** | -302.16 | -2.22 | -4.75 | -7.50 | -305.94 |

* p<0.05, ** p<0.01, *** p<0.001

**Supplementary Table S7: Regression analysis with robust standard errors of the relations between Prize Characteristics and Magnitude of Extraordinary Growth (#Incumbents). OLS regression are used in these analyses.**

|  | #Entrants | #Entrants | #Entrants | #Entrants | #Entrants |
| --- | --- | --- | --- | --- | --- |
| **Recency** | 0.0142*** |  |  |  | 0.0141*** |
|  | (0.00103) |  |  |  | (0.00103) |
| **Money (Yes/No)** |  | 0.0552*** |  | 0.0566*** | 0.0391* |
|  |  | (0.0157) |  | (0.0157) | (0.0157) |
| **Discipline/General** |  |  | 0.0810*** | 0.0824*** | 0.0876*** |
|  |  |  | (0.0179) | (0.0179) | (0.0177) |
| **Prizewinner top** | -0.0251 | -0.00691 | -0.00949 | -0.00984 | -0.0283 |
|  | (0.0168) | (0.0169) | (0.0169) | (0.0169) | (0.0168) |
| **Prize Age** | -0.0000465 | 0.00000967 | -0.00000665 | 0.0000376 | 0.0000124 |
|  | (0.0000927) | (0.0000913) | (0.0000907) | (0.0000911) | (0.0000933) |
| **Conferral times** | 0.0000135 | -0.000123 | 0.000164 | 0.000141 | 0.000275 |
|  | (0.000211) | (0.000210) | (0.000221) | (0.000220) | (0.000221) |
| **Pageviews** | 0.00000253 | 0.00000236 | 0.00000370* | 0.00000309* | 0.00000290* |
|  | (0.00000146) | (0.00000149) | (0.00000148) | (0.00000149) | (0.00000147) |
| **Discipline** | Yes | Yes | Yes | Yes | Yes |
| **Prizewinning year** | Yes | Yes | Yes | Yes | Yes |
| **Multiple recipients** | Yes | Yes | Yes | Yes | Yes |
| **Lagged variable** | Yes | Yes | Yes | Yes | Yes |
| const | 0.811*** | 0.659*** | 0.585*** | 0.568*** | 0.703*** |
|  | (0.0957) | (0.0965) | (0.0992) | (0.0996) | (0.0991) |
| **R-square** | 0.294 | 0.278 | 0.278 | 0.279 | 0.295 |
| **ΔBIC** | -279.28 | -3.10 | -9.76 | -13.51 | -289.35 |

* $p<0.05$, ** $p<0.01$, *** $p<0.001$

**Supplementary Table S8:** Regression analysis with robust standard errors of the relations between Prize Characteristics and Magnitude of Extraordinary Growth (#Entrants ). OLS regression are used in these analyses.

|  | #Disciplinay Stars working on the topic | #Disciplinay Stars working on the topic | #Disciplinay Stars working on the topic | #Disciplinay Stars working on the topic | #Disciplinay Stars working on the topic |
|---|---|---|---|---|---|
| **Recency** | 0.0146*** |  |  |  | 0.0146*** |
|  | (0.00109) |  |  |  | (0.00109) |
| **Money (Yes/No)** |  | 0.0562*** |  | 0.0582*** | 0.0401* |
|  |  | (0.0166) |  | (0.0166) | (0.0166) |
| **Discipline/General** |  |  | 0.120*** | 0.121*** | 0.127*** |
|  |  |  | (0.0192) | (0.0192) | (0.0189) |
| **Prizewinner top** | -0.0188 | -0.0000501 | -0.00399 | -0.00436 | -0.0234 |
|  | (0.0176) | (0.0177) | (0.0177) | (0.0177) | (0.0176) |
| **Prize Age** | -0.0000226 | 0.0000347 | 0.0000303 | 0.0000758 | 0.0000498 |
|  | (0.000114) | (0.000110) | (0.000108) | (0.000109) | (0.000114) |
| **Conferral times** | -0.0000554 | -0.000195 | 0.000217 | 0.000193 | 0.000331 |
|  | (0.000225) | (0.000225) | (0.000235) | (0.000235) | (0.000235) |
| **Pageviews** | 0.00000180 | 0.00000164 | 0.00000334* | 0.00000272 | 0.00000252 |
|  | (0.00000165) | (0.00000169) | (0.00000168) | (0.00000169) | (0.00000167) |
| **Discipline** | Yes | Yes | Yes | Yes | Yes |
| **Prizewinning year** | Yes | Yes | Yes | Yes | Yes |
| **Multiple recipients** | Yes | Yes | Yes | Yes | Yes |
| **Lagged variable** | Yes | Yes | Yes | Yes | Yes |
| const | 0.851*** | 0.694*** | 0.578*** | 0.560*** | 0.700*** |
|  | (0.0986) | (0.0992) | (0.101) | (0.102) | (0.101) |
| **R-square** | 0.297 | 0.281 | 0.283 | 0.283 | 0.299 |
| **ΔBIC** | -266.40 | -2.24 | -28.30 | -31.44 | -296.11 |

* p<0.05, ** p<0.01, *** p<0.001

**Supplementary Table S9:** Regression analysis with robust standard errors of the relations between Prize Characteristics and Magnitude of Extraordinary Growth (#Disciplinay Stars working on the topic ). OLS regression are used in these analyses.

|  | Productivity | Citations | Impact of Topic's Lead Scientists | #Incumbents | #Entrants | #DisciplinayStars working on the topic |
| --- | --- | --- | --- | --- | --- | --- |
| Est1 r2 | 0.321 | 0.352 | 0.111 | 0.263 | 0.262 | 0.296 |
| Est2 r2 | 0.278 | 0.363 | 0.092 | 0.283 | 0.277 | 0.294 |
| Est3 r2 | 0.309 | 0.359 | 0.105 | 0.232 | 0.265 | 0.317 |
| Est4 r2 | 0.354 | 0.383 | 0.107 | 0.228 | 0.302 | 0.260 |
| Est5 r2 | 0.304 | 0.383 | 0.117 | 0.242 | 0.322 | 0.251 |
| Est6 r2 | 0.316 | 0.337 | 0.103 | 0.296 | 0.331 | 0.281 |
| Est7 r2 | 0.307 | 0.381 | 0.106 | 0.292 | 0.277 | 0.319 |
| Est8 r2 | 0.323 | 0.336 | 0.089 | 0.198 | 0.307 | 0.274 |
| Est9 r2 | 0.273 | 0.384 | 0.130 | 0.282 | 0.258 | 0.326 |
| Est10 r2 | 0.291 | 0.382 | 0.094 | 0.251 | 0.251 | 0.282 |
| All data r2 | 0.318 | 0.375 | 0.120 | 0.267 | 0.296 | 0.300 |

**Supplementary Table S10. 10-Fold Cross-Validation for Regression Results.**

|  | Productivity | Citations | Impact of Topic's Lead Scientists | #Incumbents | #Entrants | #Disciplinay Stars working on the topic |
| --- | --- | --- | --- | --- | --- | --- |
| **No money** | 0 | 0 | 0 | 0 | 0 | 0 |
|  | (.) | (.) | (.) | (.) | (.) | (.) |
| **Low money** | 0.0451* | 0.0122 | -0.0419 | 0.0703** | 0.0157 | 0.0301 |
|  | (0.0214) | (0.0199) | (0.0216) | (0.0252) | (0.0223) | (0.0242) |
| **High Money** | 0.0517** | 0.0754*** | 0.00419 | 0.0363 | 0.0718*** | 0.0578** |
|  | (0.0196) | (0.0185) | (0.0197) | (0.0230) | (0.0209) | (0.0218) |
| **Const** | 0.328*** | 0.277*** | 0.231*** | 0.433*** | 0.305*** | 0.383*** |
|  | (0.0117) | (0.0110) | (0.0121) | (0.0139) | (0.0124) | (0.0130) |

* $p<0.05$, ** $p<0.01$, *** $p<0.001$

**Supplementary Tab S11. Robustness check of the relations between Prize Characteristics and the money of prizes. We identified the median of money amounts for prizes with money and we separate prizes into three groups. OLS regression are used in these analyses.**

|  | Description |
|---|---|
| **Recency** | How many years the prizewinner has been working on the topic before the prizewinning year. (mean=11.40, STD=9.46) |
| **Money** | This dummy variable takes on the value of 1 if prize money is associated with the prize and 0 otherwise (45% of the prizes contains money). |
| **Discipline-specific or General prize** | This dummy variable takes on the value of 1 if at least 85% of all the winners of the prize come from the same discipline and 0 otherwise (78% are discipline-specific prizes). |
| **Prizewinner top** | This dummy variable takes on the value of 1 if the prizewinner is among top 5% scientists within the topic in the prizewinning year based on citation records and 0 otherwise. (28.3% of the prizewinners are among the top 5%). |
| **Prize Age** | The age of the prize in years before the prizewinning year (mean=27.47, STD=29.94). |
| **Number of times the prize was conferred** | The number of times the prize was bestowed before the prizewinning year (mean=26.08, STD=36.14). |
| **Pageviews** | The number of wikipedia pageviews of the prize before the end of 2017. |
| **Discipline** | This categorical variable is used to define each of 19 separate disciplines. |
| **Prizewinning year** | This categorical variable is used to define the calendar year of the prizewinning event (1970-2007). |
| **Multiple recipients** | This dummy variable takes on the value of 1 if there are multiple prizewinners for the topic in the prizewinning year and 0 otherwise (4.8% of the topics have multiple prizewinners). |
| **Lagged dependent variables** | Lagged values of each growth trend at times $t$-1, $t$-2, and $t$-3 years. |

**Supplementary Tab S12. Description of Regression Variables and Their Statistics.**

**A Topic's Comparative Post-Prizewinning Growth on Six Measures – Both prizewinning topics and peer topics are NIH funded**

|  | (1) | (2) | (3) | (4) | (5) | (6) |
|---|---|---|---|---|---|---|
|  | **Productivity** | **Citations** | **Impact of Topic's Lead Scientists** | **#Incumbents** | **#Entrants** | **#Disciplinay Stars working on the topic** |
| Prizewinning ($\beta_1$) | 0.00406 | -0.000521 | 0.00212 | -0.00111 | -0.00453 | -0.00189 |
|  | (0.0590) | (0.0598) | (0.0257) | (0.0678) | (0.0624) | (0.0540) |
| Post ($\beta_2$) | 0.641*** | 1.384*** | 0.960*** | 1.048*** | 0.787*** | 0.796*** |
|  | (0.0189) | (0.0224) | (0.0132) | (0.0240) | (0.0191) | (0.0204) |
| Prizewinning * Post ($\beta_3$) | 0.132*** | 0.0896*** | 0.0856*** | 0.173*** | 0.124*** | 0.166*** |
|  | (0.0230) | (0.0269) | (0.0175) | (0.0290) | (0.0235) | (0.0246) |
| Fixed Effect Controls: |  |  |  |  |  |  |
| Discipline | Yes | Yes | Yes | Yes | Yes | Yes |
| Prizewinning Year | Yes | Yes | Yes | Yes | Yes | Yes |
| const | 3.806*** | 6.299*** | 7.098*** | 2.792*** | 4.415*** | 2.397*** |
|  | (0.150) | (0.146) | (0.0582) | (0.167) | (0.146) | (0.128) |
| N | 323,694 | 323,694 | 323,694 | 323,694 | 323,694 | 323,694 |
| R-sq | 0.213 | 0.318 | 0.327 | 0.251 | 0.227 | 0.268 |

Standard errors in parentheses. * $p<0.05$, ** $p<0.01$, *** $p<0.001$

**Supplementary Tab S13: DID analysis of a prizewinning topic's comparative post-prizewinning growth on six measures for topics that received NIH funding (both prizewinning and peer topics).** This table reported a new analysis where both the prizewinning and peer topics are NIH funded topics. Of the 2,853 NIH funded prizewinning topics, 2,569 prizewinning topics were matched with five non-prizewinning, NIH funded peer topics using DOM. This procedure ensures that 100% of the peer topics are also NIH funded topics and the subsample is balanced and meets the parallel trend requirement. By repeating the matching process in this NIH funded topic pool, we find the results are consistent. **OLS regression are used in these analyses.**

**References:**

1       Ma, Y. & Uzzi, B. Scientific prize network predicts who pushes the boundaries of science. *Proceedings of the National Academy of Sciences* **115**, 12608-12615 (2018).
2       Li, J., Yin, Y., Fortunato, S. & Wang, D. A dataset of publication records for Nobel laureates. *Sci Data* **6**, 33, doi:10.1038/s41597-019-0033-6 (2019).
3       Frank, M. R. *et al.* Toward understanding the impact of artificial intelligence on labor. *Proc Natl Acad Sci U S A* **116**, 6531-6539, doi:10.1073/pnas.1900949116 (2019).
4       Goh, K.-I. *et al.* The human disease network. *Proceedings of the National Academy of Sciences* **104**, 8685-8690 (2007).
5       Sinatra, R., Wang, D., Deville, P., Song, C. & Barabási, A.-L. J. S. Quantifying the evolution of individual scientific impact.  **354**, aaf5239 (2016).
6       Liu, L. *et al.* Hot streaks in artistic, cultural, and scientific careers. *Nature* **559**, 396 (2018).
7       Jin, G. Z., Jones, B., Lu, S. F. & Uzzi, B. The reverse Matthew effect: Consequences of retraction in scientific teams. *Review of Economics and Statistics* **101**, 492-506 (2019).
8       Lu, S. F., Zhe Jin, G., Uzzi, B. & Jones, B. The Retraction Penalty: Evidence from the Web of Science. *Nature Scientific Reports* **3** (2013).
9       Pimentel, S. D., Kelz, R. R., Silber, J. H. & Rosenbaum, P. R. Large, sparse optimal matching with refined covariate balance in an observational study of the health outcomes produced by new surgeons. *Journal of the American Statistical Association* **110**, 515-527 (2015).
10      Rosenbaum, P. R. Optimal matching for observational studies. *Journal of the American Statistical Association* **84**, 1024-1032 (1989).
11      Rosenbaum, P. R. Modern Algorithms for Matching in Observational Studies. *Annual Review of Statistics and Its Application* **7** (2019).
12      Stuart, E. A. Matching methods for causal inference: A review and a look forward. *Statistical science: a review journal of the Institute of Mathematical Statistics* **25**, 1 (2010).
13      Zubizarreta, J. R. Using mixed integer programming for matching in an observational study of kidney failure after surgery. *Journal of the American Statistical Association* **107**, 1360-1371 (2012).
14      Kuhn, T. S. *The Structure of Scientific Revolutions*.  (University of Chicago, 1970).
15      Jones, B. F. The burden of knowledge and the 'Death of the Renaissance Man': Is innovation getting harder? *Review of Economic Studies* (2008).